\documentclass[onecolumn,sort&compress,numbers]{els-mrw} 

\usepackage{amsmath,amssymb,amsfonts,amsthm,makeidx,graphicx}
\usepackage{txfonts}
\usepackage{helvet}
\usepackage{hyperref}

\begin{document}


\chapter{Lattice QCD calculations of hadron spectroscopy  }\label{chap1}

\author[1,2]{ Sasa Prelovsek} 

\address[1]{\orgname{University of Ljubljana }, \orgdiv{ Faculty of Mathematics and Physics}, \orgaddress{Jadranska 19, Slovenia }} 
\address[2]{\orgname{ Jozef Stefan Institute }, \orgdiv{  Department of Theoretical Physics}, \orgaddress{Jamova 39, Slovenia }}

\articletag{Chapter Article tagline: update of previous edition, reprint.}

\maketitle

\begin{abstract}[Abstract]
This chapter provides a pedagogical introduction to theoretical studies of hadrons based on the fundamental theory of strong interactions -  Quantum ChromoDynamics. A perturbative expansion in the strong coupling is not applicable at hadronic energy scales.  Lattice Quantum Chromodynamics is the formulation of the fundamental theory on a discrete space-time grid, which enables first-principles, systematically improvable, numerical simulations of strong interaction physics.  This chapter explains how the masses of strongly stable and strongly decaying hadrons are determined. The strongly decaying hadrons have to be inferred from the corresponding scattering processes. Therefore, one of the main aims is to describe how the scattering amplitudes are extracted from a lattice simulation.  The examples of spectra, widths, and scattering amplitudes are shown for conventional as well as exotic hadrons. 
\end{abstract}

\begin{keywords}
 	hadrons\sep spectroscopy\sep QCD\sep lattice QCD \sep scattering
\end{keywords}

\section*{Objective}
The objective is to learn how the following physics problems have been or can be addressed: 
\begin{itemize}
	\item  Masses of proton, neutron, and other hadrons are not just experimentally measurable quantities, but they have been theoretically determined from the fundamental theory of the strong interactions - QCD.  
	\item   Masses and lifetimes have been reliably calculated for a number of hadronic resonances,  which strongly decay via one channel.
	\item  Most of the conventional and exotic hadrons strongly decay to several final states, and some of these have already been addressed ab initio. The presented approach makes it clear why this is a challenging task.  
	\item This is not a review, but aims at a pedagogical and rather self-contained introduction, with few examples chosen for pedagogical purposes.  Review articles, for example \cite{Brambilla:2019esw,Bicudo:2022cqi,Bulava:2022ovd,Briceno:2017max,Hanlon:2024fjd},  need to be consulted for references to many other interesting studies and to grasp the current state of the art. 
	\end{itemize}

\section{Introduction}\label{intro}

 Strong interactions are the strongest among all four fundamental interactions and are responsible for binding the quarks to color-neutral hadrons. Conventional hadrons have minimal valence quark content $\bar q_1q_2$ (mesons) and $q_1q_2q_3$ (baryons).  In the past two decades, around thirty candidates for exotic hadrons with minimal valence quark content  $\bar q_1\bar q_2q_3q_4$ (tetraquarks), $\bar q_1 q_2 q_3 q_4 q_5$ (pentaquarks), and $\bar q_1Gq_2$ (hybrids) have been discovered in experiments \cite{pdg2024}.  The binding mechanisms responsible for the existence of these multi-quark states represent an important open question. 
 
 The main aim of hadron spectroscopy is to determine hadron masses and improve our understanding of emerging mass patterns. One of the main challenges is that most of the known hadrons, and in particular all the experimentally observed exotic hadrons,  are hadronic resonances. These are metastable states that decay quickly to lighter hadrons via the strong interaction, either only to one final state  $R\to H_1H_2$ or to several final states $R\to H_1H_2,~  H_1'H_2',..$.  An important aim for resonances is to determine their decay width $\Gamma$ or equivalently the lifetime $\tau=\hbar/\Gamma$,   branching ratios $Br^i=\Gamma^i/\Gamma$ for various final states, and lineshapes  $d\Gamma/dm_{H_1H_2}$ as a function of invariant masses $m_{H_1H_2}$. 
 
\begin{figure}[htb]
	\centering
	\includegraphics[width=0.8\textwidth]{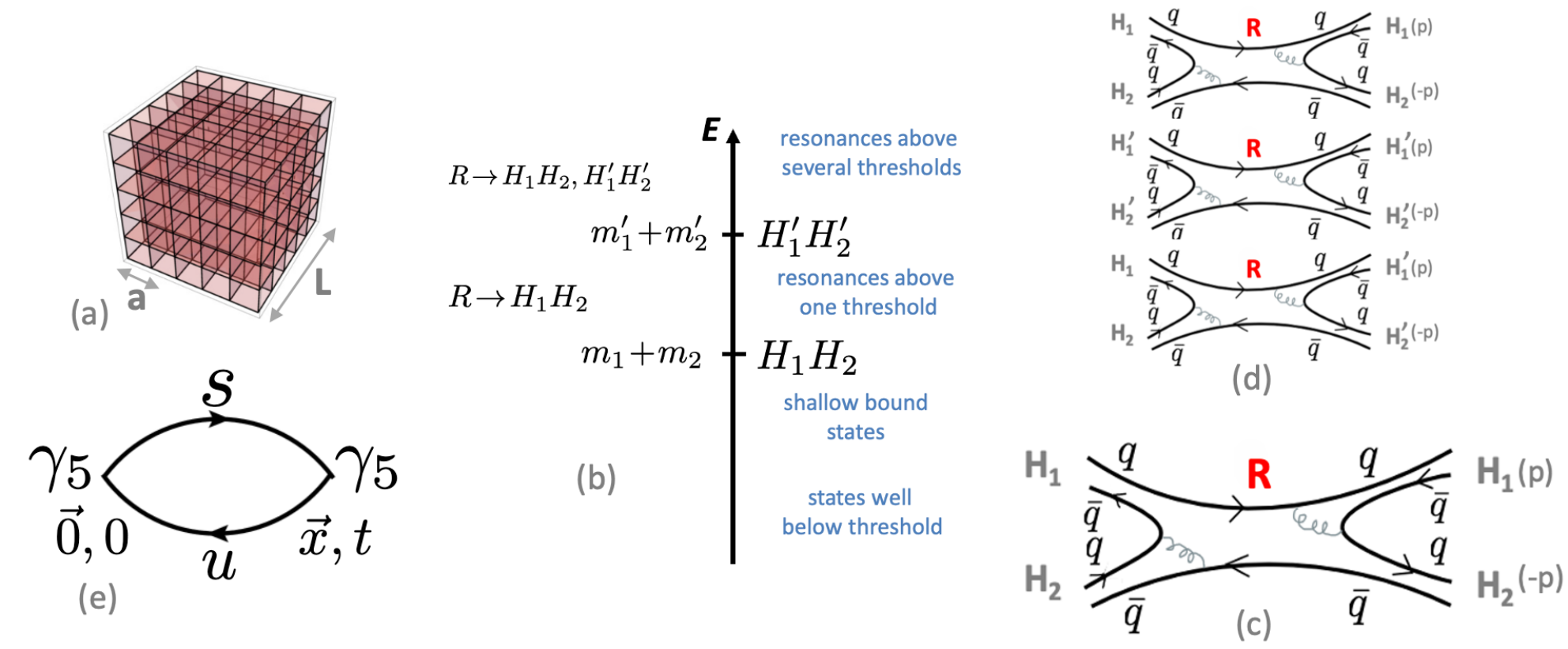}\quad
	\caption{(a) Lattice box; (b) The difficulty of a lattice study increases for states that lie above the strong decay threshold and especially above several thresholds; (c)  One-channel scattering;   (d)  Scattering for two coupled channels;   (e) Kaon correlator. }
	\label{fig:lattice}
\end{figure}

 This chapter aims at a pedagogical introduction on how to theoretically determine the physics observables listed in the previous paragraph from first-principle lattice QCD, and provide a few pedagogical examples. The only interaction at play will be the strong interactions described with ${\cal L}_{QCD}$, while electro-weak interactions will be neglected.  Let us first briefly review how difficult it is to study a given hadron, where the  difficulty increases 
 from bottom to top  as illustrated in   Figure \ref{fig:lattice}(b): 
 \begin{itemize}
\item  Hadrons that are stable with respect to strong decay and lie significantly below the lowest decay threshold are the most straightforward. 
  \item Hadrons that can decay via single channel $R\to H_1H_2$ or hadrons that reside slightly below threshold $H_1H_2$ have to be inferred from the scattering $H_1H_2\to R\to  H_1H_2$ sketched in   Figure \ref{fig:lattice}(c).   This already presents a more challenging theoretical problem since the scattering amplitude $T(E_{cm})$  has to be extracted, where  $E_{cm}$ is the center-of-momentum energy. The hadronic states correspond to the pole singularities of  $T(E_{cm})$  in the complex $E_{cm}$ plane. The pole position $E_{cm}^p=m-\frac{i}{2}\Gamma$ is related to hadron mass and width, while   $|T(E_{cm})|^2$ is related to the experimental lineshape. The ab initio theory study of the scattering amplitudes is, therefore, a prime subject of this chapter. 
    \item The hadrons residing above two or more thresholds and decaying via several decay channels represent an even more challenging problem. An example of a resonance that decays via channels $R\to H_1H_2$ (channel $a$) and  $R\to H_1'H_2'$ (channel $b$) is sketched in  Figure \ref{fig:lattice}(d).     The energy dependence of the $2\times 2$ scattering matrix $T_{ij}$  ($i,j=a,b$) with elements $T_{aa}$, $T_{ab}$ and $T_{bb}$ are of prime theoretical and experimental interest in this case. All elements of the scattering matrix $T_{ij}$ have the pole at the energy where $T^{-1}\propto 1/\det(T)$ vanishes, and the pole position $E^p_{cm}=m-\tfrac{i}{2}\Gamma$ is again related to the mass and the width of the state.       
      \item  The hadrons that can strongly decay also to the final states with more than two hadrons are even more challenging, and their study goes beyond this chapter. The past few years have evidenced impressive progress on the formalisms and on actual numerical lattice QCD results for three-hadron scattering amplitudes from the lattice.   
   \end{itemize}

   \vspace{0.2cm}
  
 The experimental determination of $\rho$ resonance parameters by measuring the cross-section $\sigma (E_{cm})$ for $\pi\pi\to\rho\to\pi\pi$ scattering follows a similar strategy as described above. However,  the experimental determination of resonance parameters for many other resonances is somewhat different since many hadrons ($D$, $B$,..) quickly decay via the electro-weak interaction. In these cases, the initial state in Figure \ref{fig:lattice}(c) can not be realized in an experiment, and resonances are produced via various other production mechanisms. 
 The experimental observables depend on the underlying scattering amplitude $H_1H_2\to R\to H_1H_2$ also in this case, which calls for their theoretical determination.

    \vspace{0.2cm}
     
 The hadrons emerge from the strong interaction between quarks at energy scales where non-perturbative effects dominate. Consequently, a perturbative expansion in the QCD gauge coupling $g_s$ is inapplicable. Lattice QCD is a widely used non-perturbative method with systematically improvable uncertainties. It is based directly on the QCD Lagrangian ${\cal L}_{QCD}(m_q,g_s)$, where the only free parameters are quark masses $m_q$ and the strong coupling $g_s$.  The spectroscopic properties of hadrons are determined from correlation functions $C$. Lattice QCD is based on the Feynman path integral approach, where the correlation function $C$ is   as an expectation value of correlation functions ${\cal  C^\prime}(G,q)$ obtained by functional integration over all gluon  $G^\mu_a(x)$ and quark $q_a(x)$ fields
 \begin{equation}
 \label{path-integral}
  C =\int {\cal D}G \int {\cal D}q \int {\cal D}\bar q~{\cal  C^\prime}(G,q)~e^{-S_E(q,G)/\hbar}, \quad {\cal L}_{QCD}={\cal L}_{QCD}(m_q,g_s)~,
\end{equation}
where ${\cal C^\prime }(G,q)$   is the correlation function for a given configuration of gluon and quark fields. 
  Each field configuration is weighted by the factor $e^{iS/\hbar}$ in the case of the Minkowski space-time. Lattice simulations employ Euclidean time $t=it_M$, and the corresponding weight   $e^{-S_E/\hbar}$ is real.    
These path integrals are evaluated via the numerical evaluation on a discretized and Euclidean space-time of finite volume $N_L^3N_T a^4$  sketched in Figure  \ref{fig:lattice}(a) \cite{Gattringer:2010zz}. Reliable physics predictions should be based on simulations at several lattice spacings $a$ and several spatial volumes $L=aN_L$, followed by the continuum and infinite volume extrapolations $a\to 0$ and $L\to \infty$. Lattice QCD can explore how the properties of hadrons depend on the quark masses of each quark flavor. This is a very valuable handle used to investigate the binding mechanisms that are responsible for their existence.

  Practically all physics observables mentioned above   can be calculated from eigen-energies $E_n$  of the QCD Hamiltonian   $H_{QCD}|n\rangle = E_n | n \rangle$ on the finite lattice, as argued in the reminder of this chapter.   This applies also to the scattering amplitude $T(E)$ derived in Section \ref{sec:one-channel}. The next section introduces how to determine eigen-energies from the so-called two-point correlation functions $C$, which are, in turn, evaluated via the path integral above.

 \section{Eigen-energies from a lattice simulation }\label{sec:En}

Spectroscopic information for a hadron that resides below, near, or above threshold is commonly extracted from the discrete set of energies $E_n$ of QCD eigenstates $|n\rangle$ on a finite-volume lattice. These are determined from the two-point correlation functions $C_{ij}(t)=\langle \Omega| \hat O_i(t)\hat O_j^\dagger (0)|\Omega\rangle$, where $|\Omega\rangle$ is the vacuum, $O_j^\dagger $ creates a hadron system with quantum numbers of interest, and $O_i^\dagger $ annihilates it after time $t$. 

The operators $O_i(q(x),G(x))$ are built out of quark and gluon field operators and should have the same quantum numbers as the system of interest: these are typically flavor quantum numbers,  total momentum  $\vec P$ and the quantum numbers related to parity and spin\footnote{A lattice with a finite spatial extent has a reduced symmetry compared to an infinite-volume continuum. This means that instead of $J$, the relevant quantum number is an irreducible representation of the appropriate point group, for example, a cubic group in the case of momentum zero.}.  Care must be taken to ensure that operators have appropriate structures to effectively overlap with the eigenstates of interest.  They create/annihilate the eigenstates of interest, as well as, in principle, all other eigenstates $|n\rangle$ with a given quantum number. Examples of operators in the meson sector   with conventional or exotic quark content are 
\begin{equation}
\label{operators}
O(t): ~ \sum_{\vec x}\! e^{i\vec P\vec x} ~\bar q(\vec x,t) \Gamma q(\vec x,t) ~, \  \sum_{\vec x}\!  e^{i\vec P\vec x}~(\bar q \Gamma_1 q) ~(\bar q \Gamma_2 q)~,\  \sum_{\vec x}\! e^{i\vec P\vec x} ~ \epsilon_{abc} [q^{bT} \Gamma_1 q^c]~  \epsilon_{ade} [\bar q^d \Gamma_2 \bar q^{eT}]~ , \ \sum_{\vec x_1} \!e^{i\vec p_1\vec x_1} \bar q(x_1) \Gamma_1 q(x_1) ~ \sum_{\vec x_2} \! e^{i\vec p_2\vec x_2}  \bar q(x_2) \Gamma_2 q(x_2)~,\nonumber
\end{equation}
where quarks fields $q$ carry various flavors and apply at the same $x=(\vec x,t)$, except in the last operator. There, each meson is separately projected to a given momentum with $\vec p_1+\vec p_2= \vec P$ and $t=x_{1,2}^0$.   
 Typical simulations in Sections \ref{sec:one-channel} and  \ref{sec:coupled} employ around $N=5-50$ operators with a given quantum number at the source and the sink, and the resulting correlation matrix $C_{ij}$ is of size $N\times N$. 

 The dependence of the correlation function  on  $E_n$ is obtained upon insertion of the complete set $I=\sum_n |n\rangle \langle n|$ of QCD eigenstates $|n\rangle$ on a finite volume of the lattice
\begin{align}
\label{C2pt}
C_{ij}(t)=\langle \Omega | \hat O_i(t)\hat O_j^\dagger (0)|\Omega\rangle=\sum_n \langle \Omega | e^{\hat H t}  \hat O_i(0)e^{-\hat H t}|n\rangle \langle n|\hat O_j^\dagger (0)|\Omega\rangle
= \sum_n \langle \Omega |  \hat O_i(0)|n\rangle e^{-E_n t} \langle n|\hat O_j^\dagger (0)|\Omega\rangle=\sum_{n=1}^\infty Z_i^{(n)} Z_j^{(n)*}e^{-E_n t},\quad Z_i^{(n)}\equiv \langle  \Omega |  \hat O_i|n\rangle~.
\end{align}
Here the eigenvalue equation $\hat H|n\rangle = E_n | n \rangle$ and the evolution of operators $ \hat O_i(t_M) =e^{i\hat H t_M}  \hat O_i(0)e^{-i\hat H t_M} =e^{\hat H t}  \hat O_i(0)e^{-\hat H t}$ on Euclidean time $t$ were employed. The overlap $Z$ of a given eigenstate and operator provides valuable qualitative information.

As an example,  the correlation function for a system with kaon quantum numbers and total momentum $\vec P$  is  calculated with the path integration (\ref{path-integral}) using  the point-like creation operator   and 
annihilation operator  $\sum_{\vec x} e^{i\vec P\vec x}\bar u(\vec x,t)\gamma_5 s(\vec x,t)$  as illustrated in Figure \ref{fig:lattice}(e) 
\begin{equation}
\label{K}
C(t) =\sum_{\vec x} e^{i\vec P\vec x}    \langle \Omega | \bar u(\vec x,t)\gamma_5 s(\vec x,t)\bar s(\vec 0,0)\gamma_5 u(\vec 0,0) |  \Omega \rangle \propto - \int {\cal D}G ~[\mathrm{det} (D)]^{N_f}~ e^{-S_G}~{\cal  C^{ \prime \prime}(G)~,\quad {\cal C^{ \prime \prime}}(G)=}\mathrm{Tr}[\gamma_5 D^{-1}_{s,~\vec x,t\leftarrow \vec 0,0}\gamma_5 D^{-1}_{u,~\vec 0,0\leftarrow \vec x,t}]~.
\end{equation}
The right-hand side follows after the path integration over the quark fields in $C \propto \int \!\! {\cal D}G{\cal D}q{\cal D}\bar q~{\cal  C^\prime }(G,q)~e^{-S_G-\bar q D(G) q}$.  According to the Wick theorem,  the path integral over quark fields renders the product of quark propagators $D^{-1}$  and $[\det(D)]^{N_f}$ for a theory with $N_f$ degenerate quarks  (see Section 5.1 of \cite{Gattringer:2010zz}). For kaon correlation function,  the integrand contains the kaon correlator ${\cal C^{ \prime \prime}}(G)$ on a given gluon field configuration, which is    a traced product of $s$-quark and $u$-quark propagators between space-time points $(\vec 0,0)$ and $(\vec x,t)$ and appropriate $\Gamma$ matrices, as  illustrated in  Figure \ref{fig:lattice}e. The quark propagator $D(G)^{-1}$ on a given background gluon field $G$ is obtained by inverting the Dirac matrix $D(G)$ on the lattice,  which is a discretized and Euclidean version of the continuum operator $D_q(G)=i\gamma_\mu(\partial^\mu +i g_sG_a^\mu T^a) -m_q$.    $D(G)$ is a large matrix of typical size $ {\cal N}\times {\cal N}$ with ${\cal N}=N_L^3\cdot N_T\cdot 3\cdot 4$, where $3$ and $4$ are dimensions of color and spinor   spaces, respectively. Therefore, its inverse and determinant are among the numerically intensive parts of the calculation.  
  The final correlator is   obtained by summing the correlator $\quad {\cal C^{ \prime \prime}}(G)=\mathrm{Tr}[..]$ over  the gluon field configurations $G$, each weighted by  $ [\mathrm{det} (D)]^{N_f}~ e^{-S_G}$ \cite{Gattringer:2010zz}. In practice, the ensemble of gauge configurations is prepared such that a given configuration is generated with a probability proportional to $ [\mathrm{det} (D)]^{N_f}~ e^{-S_G}$.

The study of a hadron with mass $m$ requires the extraction of all eigenstates with the same quantum numbers and energies $E_{cm}\lesssim m$. In order to extract the energies $E_n$ and overlaps $Z$ via relation (\ref{C2pt}), the $N\times N$ correlation matrix $C_{ij}(t)$ with $N$ operators is calculated. Then  the eigenvalue problem  $C(t)u^{(n)}(t)=\lambda^{(n)}(t)C(t_0)u^{(n)}(t)$ is solved for times $t$ larger then a fixed reference time $t_0$   \cite{Luscher:1990ck}. Let us show that the time-dependence of the eigenvalues $\lambda^{(n)}$  renders $E_n$ via $\lambda^{(n)}(t)\propto e^{-E_nt}$ at large $t$.  Suppose $t_0$ is large enough that only the lowest $N$ eigenstates contribute to the correlation function (\ref{C2pt}) in the time region between $t_0$ and $t$, i.e. 
$C_{ij}(t)=\sum_{m=1}^N Z_i^{(m)} Z_j^{(m)*}e^{-E_m t}$. Then the  vectors $u^{(n)}$   satisfying 
$\sum_{i=1}^N u_i^{(n)} Z_i^{(m)*}=\delta_{nm}$ can be found, which implies that $C_{ij}(t)u_j^{(n)}=\sum_{m=1}^N Z_i^{(m)} Z_j^{(m)*}e^{-E_m t}u_j^{(n)}=Z_i^{(n)*}e^{-E_n t}$  decays as a single-exponential. The eigenvalue, therefore, also decays as a single exponential $\lambda^{(n)}(t)=e^{-E_n(t-t_0)}$ with the energy of interest.  In practice, $E_n$ are determined from one-exponential $\lambda^{(n)}(t)=Ae^{-E_nt}$ or two-exponential  $\lambda^{(n)}(t)=Ae^{-E_nt}+A^\prime e^{-E_n^\prime t}$  fits at large enough $t$ since typically  more than $N$ eigenstates contribute   in the time region between $t_0$ and $t$.  The described and widely used generalized eigenvalue approach (GEVP) also allows the determination of the overlaps $Z_{i}^{(n)}$.

\section{Strongly-stable hadrons well below threshold}\label{sec:stable}

\begin{figure}[htb]
	\centering
	\includegraphics[width=0.47\textwidth]{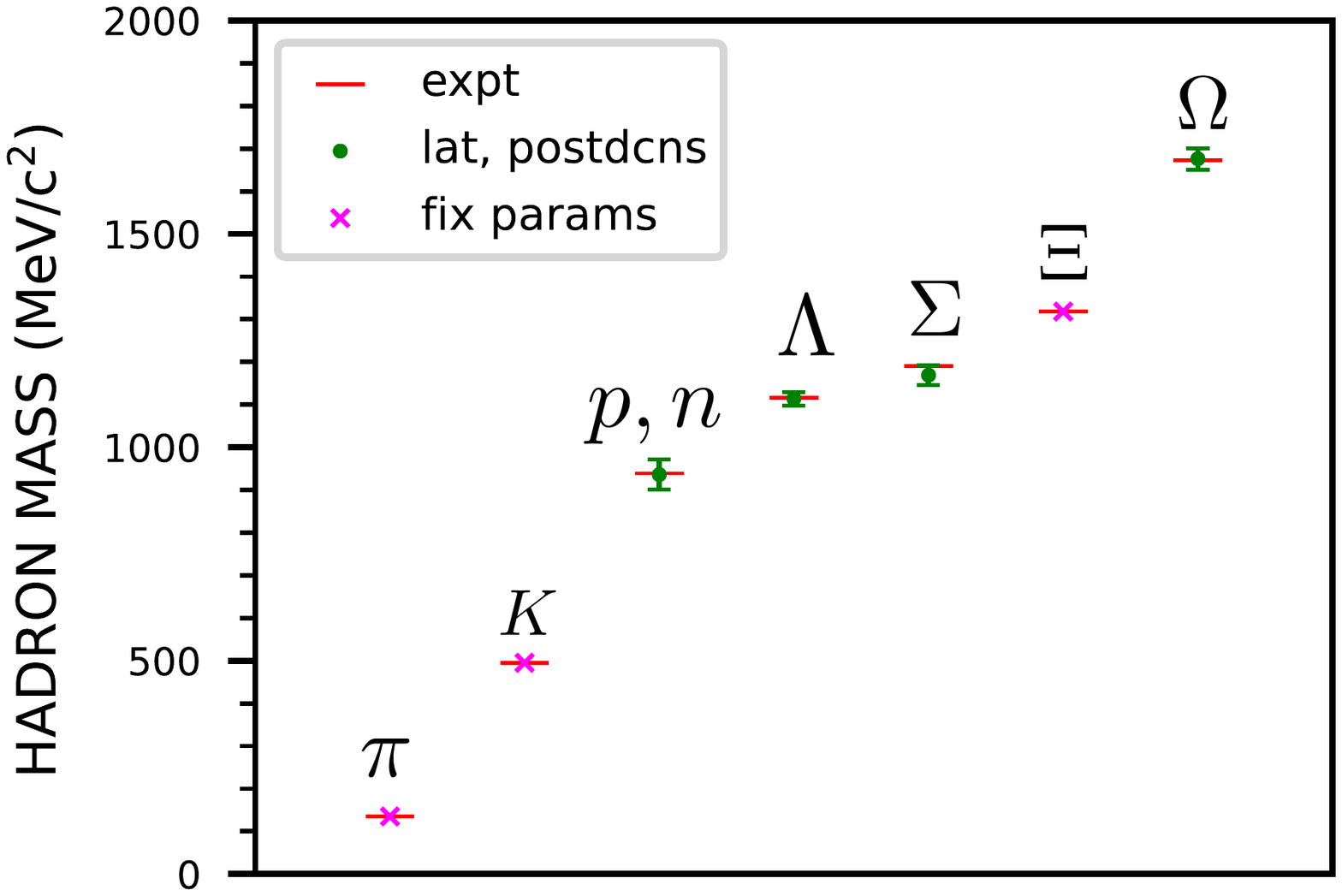}\quad
	\includegraphics[width=0.47\textwidth]{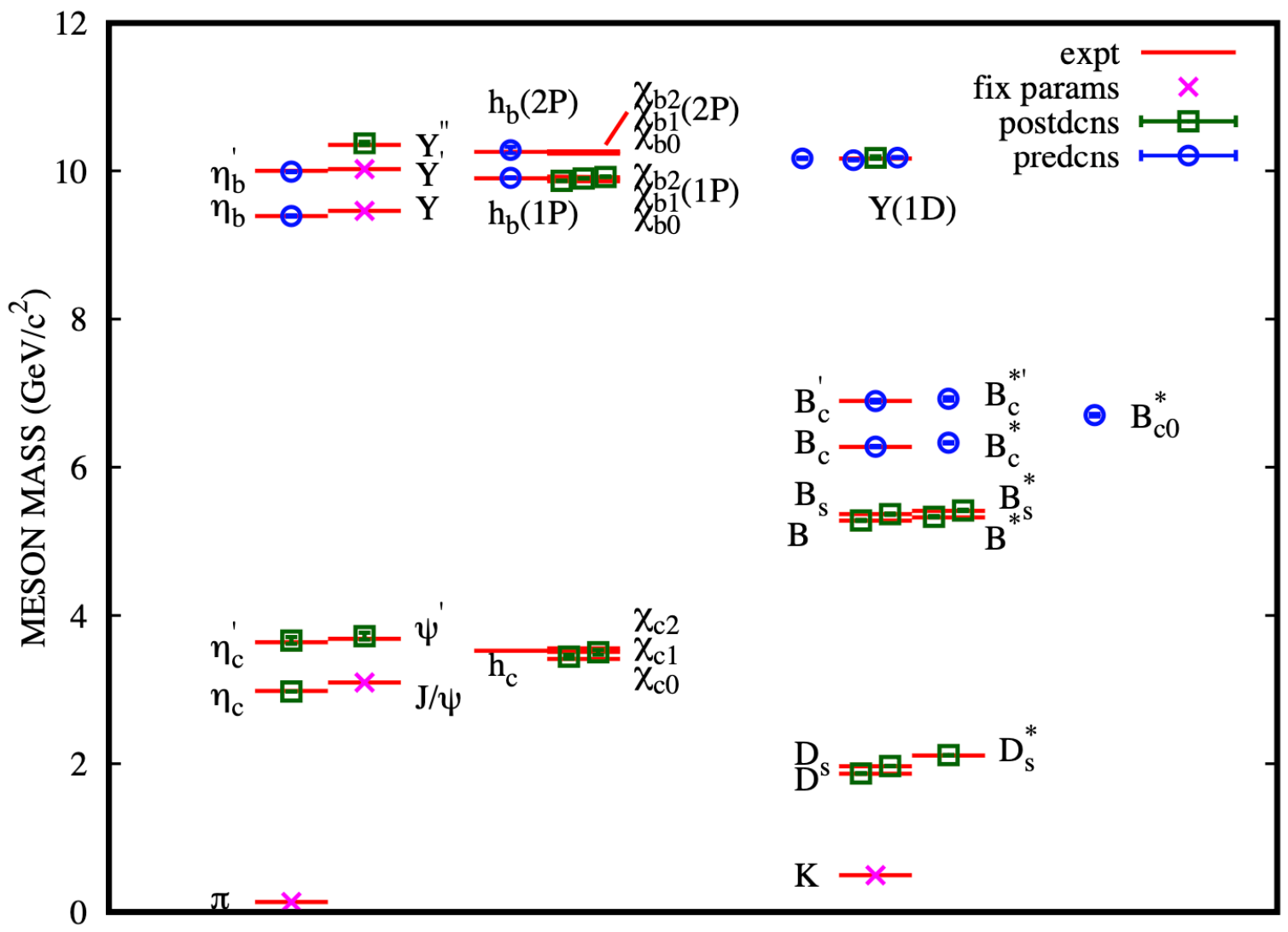}
	\caption{The masses of strongly-stable hadrons from BMW \cite{BMW:2008jgk} and HPQCD (update of \cite{Dowdall:2012ab}) simulations at physical masses of quarks in the isospin limit $m_u=m_d$.     The experimental masses indicated by crosses were used to fix the quark masses or the lattice spacing. }
	\label{fig:stable}
\end{figure}

Among the vast number of hadrons, only a handful do not decay via the strong interaction.  In the absence of electroweak interactions, the stable hadrons are the lowest-lying states of a given flavor content: $\pi$, $K$, $D$, $B^{(*)}$, $B_c^{(*)}$, $p$, $n$, $\Lambda$, $\Lambda_c$, $\Xi_{cc}$, etc.
Their masses are obtained from ground state energies as  $m = E_1 ( \vec P\!=\!\vec 0)$ after these energies are extrapolated to zero lattice spacing and large volume.  The ground state energies are calculated from the two-point correlation functions (\ref{C2pt}).  A number of lattice QCD simulations have already determined masses of such hadrons with sub-percent statistical precision and with all systematic uncertainties quantified or removed.  The lattice results for the masses of proton, neutron, and other conventional hadrons in Figure \ref{fig:stable}  \cite{BMW:2008jgk,Dowdall:2012ab} show good agreement with the experiment.  

\begin{figure}[htb]
	\centering
	\includegraphics[width=0.49\textwidth]{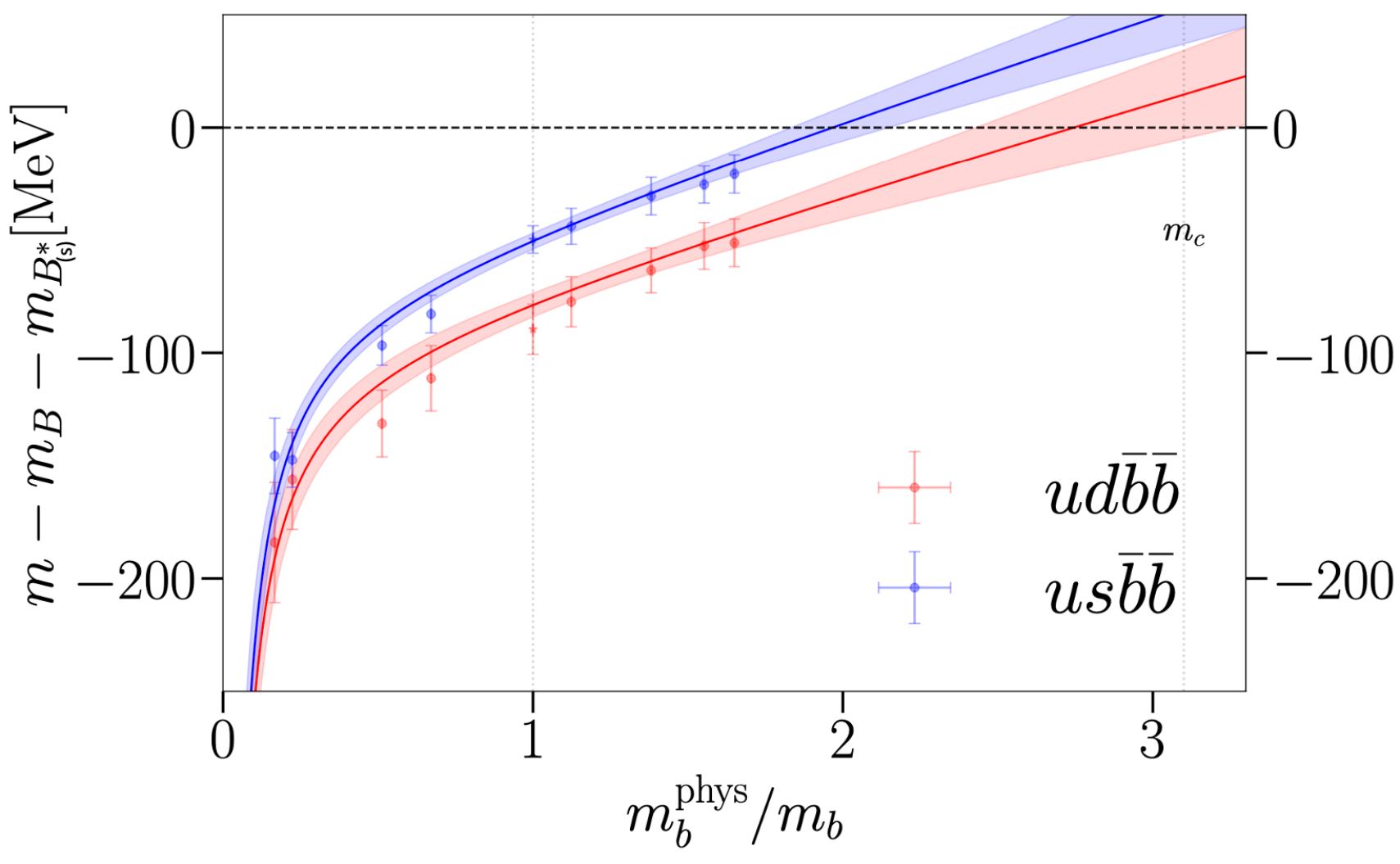}\quad \includegraphics[width=0.49\textwidth]{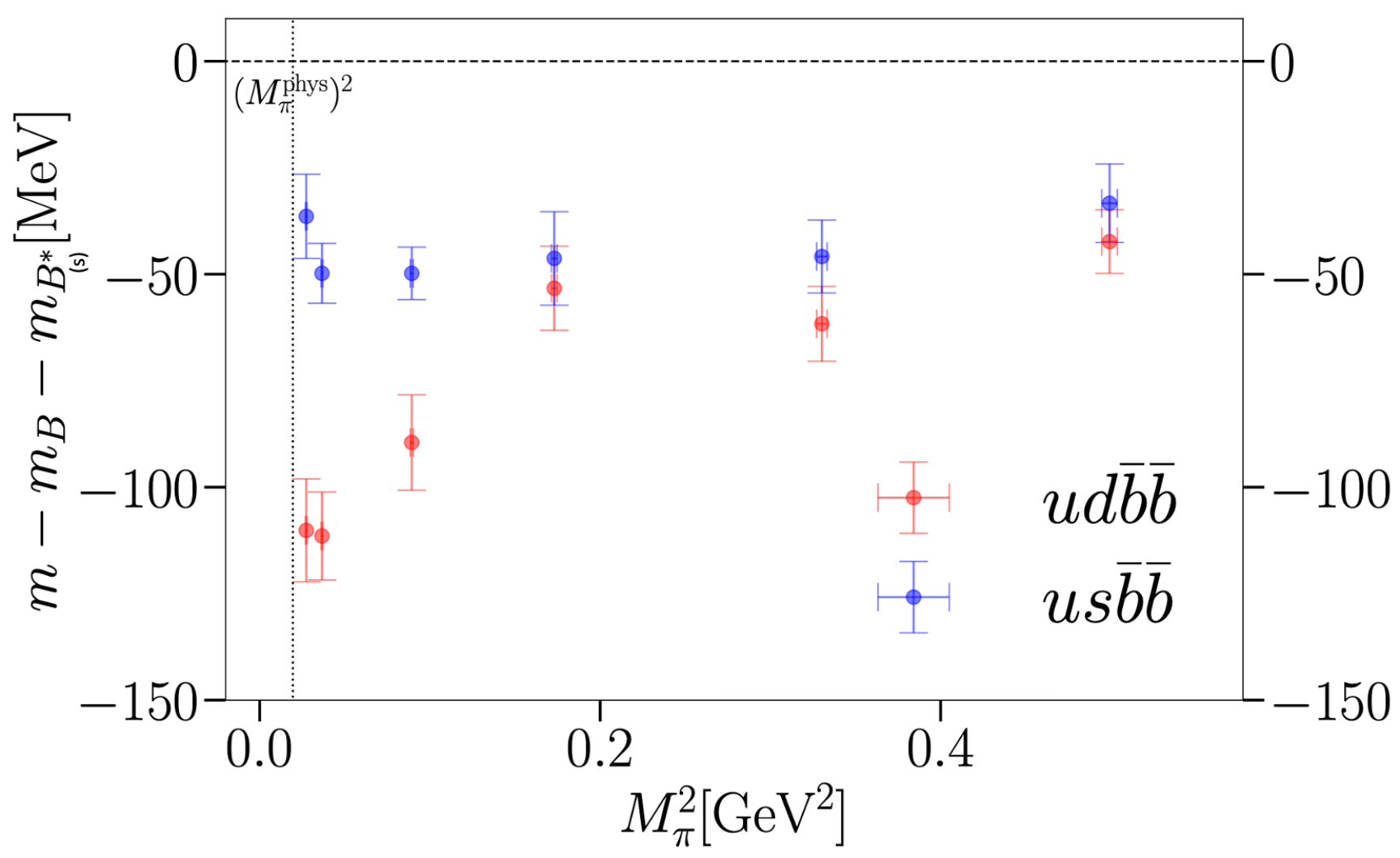} 
		\caption{The binding energies $m-m_B-m_{B_{(s)}}^*$ of tetraquarks $ \bar b\bar b ud$  and $\bar  b\bar b u s$  with $J^P=1^+$ 
	for various quark masses $m_b$ and $m_{u/d}$ from the lattice simulation 	 \cite{Colquhoun:2024jzh}. Here $m$ is the mass of a tetraquark, and $BB_{(s)}^*$ is the lowest strong decay threshold.}
	\label{fig:Tbb}
\end{figure}

Most of the exotic hadrons can strongly decay. Two exceptions are tetraquarks  $\bar b \bar b u d$ and $\bar b\bar bus$  with $J^P=1^+$ that are expected to lie significantly below $BB_{(s)}^*$ thresholds.  Their masses have been determined  from the ground state energy, and 
their binding energies $m-m_B-m_{B^*_{(s)}}$     increase with increasing $m_b$ and decreasing $m_{u/d}$,  as shown in Figure \ref{fig:Tbb} \cite{Colquhoun:2024jzh}. 

\section{Hadrons from one-channel scattering}\label{sec:one-channel}

\begin{figure}[htb]
	\centering
	\includegraphics[width=0.35\textwidth]{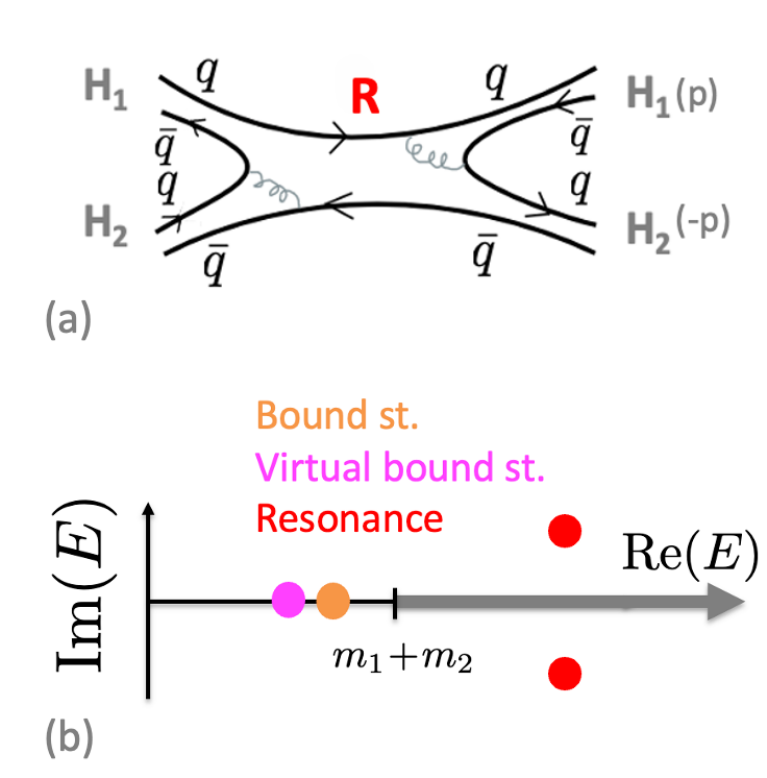}\qquad \includegraphics[width=0.55\textwidth]{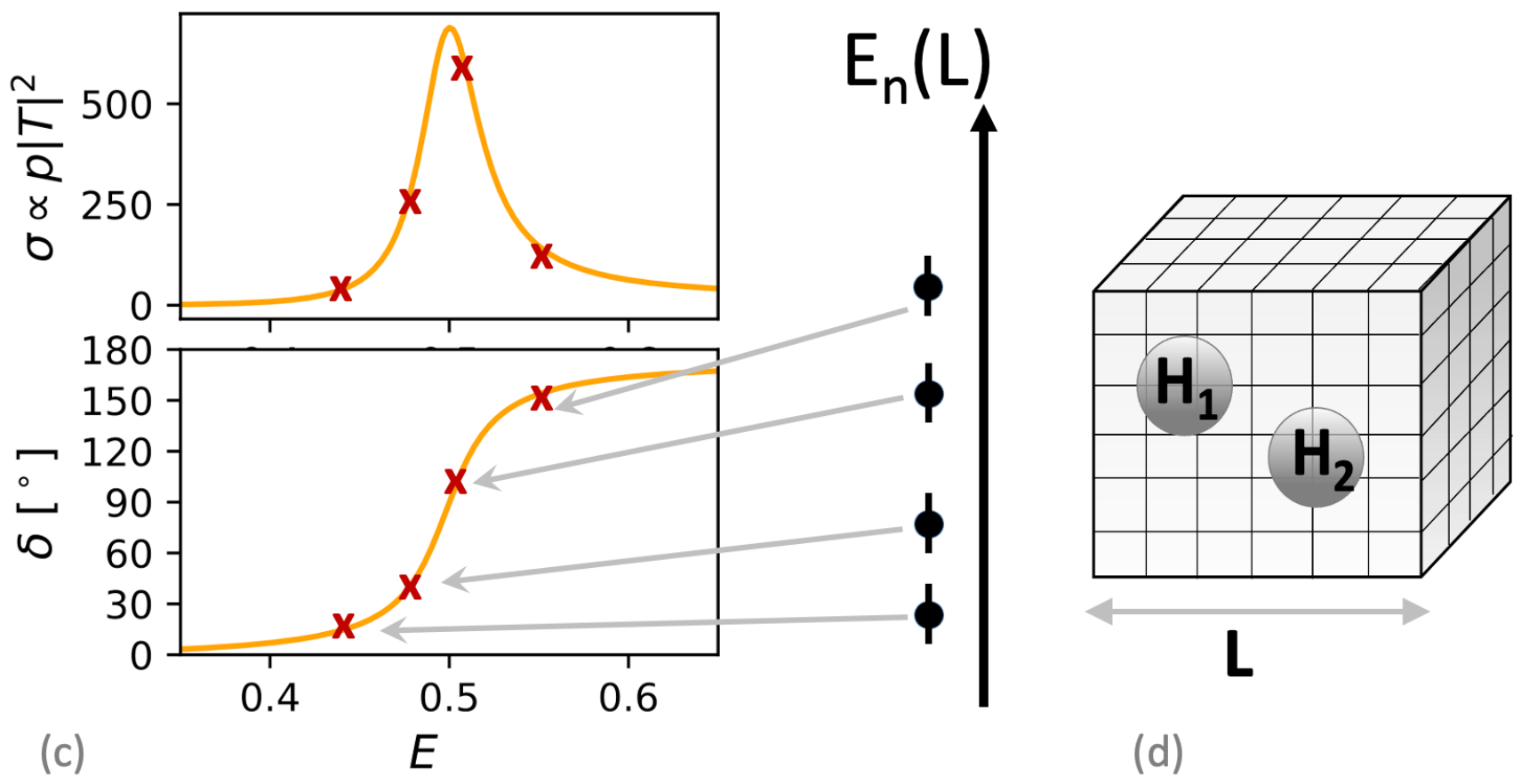}
		\caption{One-channel scattering:  (a) scattering; (b)  poles of the scattering amplitude $T$ in the complex energy-plane; (c) Extraction of the scattering amplitude from eigen-energies on the lattice via the L\"uscher's method; (d) two hadrons in a box and the corresponding eigen-energies.   }
	\label{fig:one-channel-sketch}
\end{figure}

Most hadrons are {\it hadronic resonances} that decay via the strong interaction. In particular, nearly all experimentally discovered exotic hadrons decay strongly. Such states are not asymptotic states of QCD, and their existence has to be inferred from the amplitude $T(E)$ for the scattering illustrated in Figure \ref{fig:one-channel-sketch}(a).  Let us consider the simplest case of a single-channel scattering that is completely dominated by the partial wave $l$. The conservation of probability implies that  $S$-matrix can be expressed in terms of a phase shift $\delta_l(p)$
\begin{equation}
\label{T}
S_l(p)=e^{2i\delta_l(p)}=1+2i \tfrac{p}{8 \pi E} T_l(p) \quad \to \quad  T_l(p)=8\pi E \frac{1}{p\cot\delta_l(p) - ip}~,
\end{equation}
 where $p=|\vec p|$ is the on-shell three momentum of the scattering particles in the center-of-momentum  frame\footnote{Various normalizations of $T$ are used in the literature. Here $T$ is ${\cal M}$  of Ref.  \cite{Kim:2005gf}, adapted by a factor of 2 suited for the scattering of non-identical particles in Section  \ref{sec:luscher-QFT}.   }.
 The cross-section is proportional to $|T_l|^2$, so the energy dependence of  $T_l$ is the main quantity of interest. 
 
 A resonance corresponds to a pole of the scattering amplitude $T_l$ away from the real axes, as sketched in Figure \ref{fig:one-channel-sketch}(b). Bound
state and virtual state correspond to a pole at $E=(p^2+m_1^2)^{1/2}+(p^2+m_2^2)^{1/2}<m_1+m_2$ below threshold,   therefore at real negative $p^2$. 
The bound state pole at positive imaginary $p=i|p|$  is related to an asymptotic bound state with decreasing wave-function $e^{ipr}=e^{-|p|r}$ outside the region of interaction. The virtual state pole  $p = -i|p|$   is not related to an asymptotic state as its wave function increases as $e^{|p|r}$ outside the region of interaction. Both bound and virtual states lead to a significant enhancement of the scattering rate above the threshold if the poles are located only slightly below the threshold.  
 
\subsection{Relation between finite-volume eigen-energies and infinite-volume scattering amplitude}
\label{sec:luscher}

The most widely applied and rigorous method to extract the scattering matrix from ab-initio lattice QCD simulations is the so-called {\it L\"uscher formalism} \cite{Luscher:1990ux,Kim:2005gf}. In order to determine the amplitude for  $H_1H_2$ scattering in a channel with certain quantum numbers, the eigen-energies of a system with these quantum numbers have to be calculated from lattice QCD. The  L\"uscher formalism rigorously relates the discrete eigenenergy $E_{cm}$ of the system in a finite volume, sketched in Figure \ref{fig:one-channel-sketch}(c),  with the infinite volume scattering matrix $T(E_{cm})$ at the same  $E_{cm}$
\begin{equation}
E_{cm}(L) \leftrightarrow  T(E_{cm})~.  
\end{equation}
An explicit example of this relation is derived below and provided in Eq. (\ref{luscher}) for the simplest case of scattering with the periodic boundary conditions, $l=0$, and total momentum zero. The generalized relation for the scattering of particles with arbitrary spin and partial wave is provided in Eq. (22) of \cite{Briceno:2014oea}. 
Lattice studies aim to extract a large number of eigen-energies in order to constrain the scattering amplitude at various energies $E_{cm}$. Several lattice sizes $L$ and various total momenta $\vec P$ are employed for this purpose.

\paragraph{\bf One-dimensional quantum mechanics}
 
\begin{figure}[htb!]
	\centering
	\includegraphics[width=0.95\textwidth]{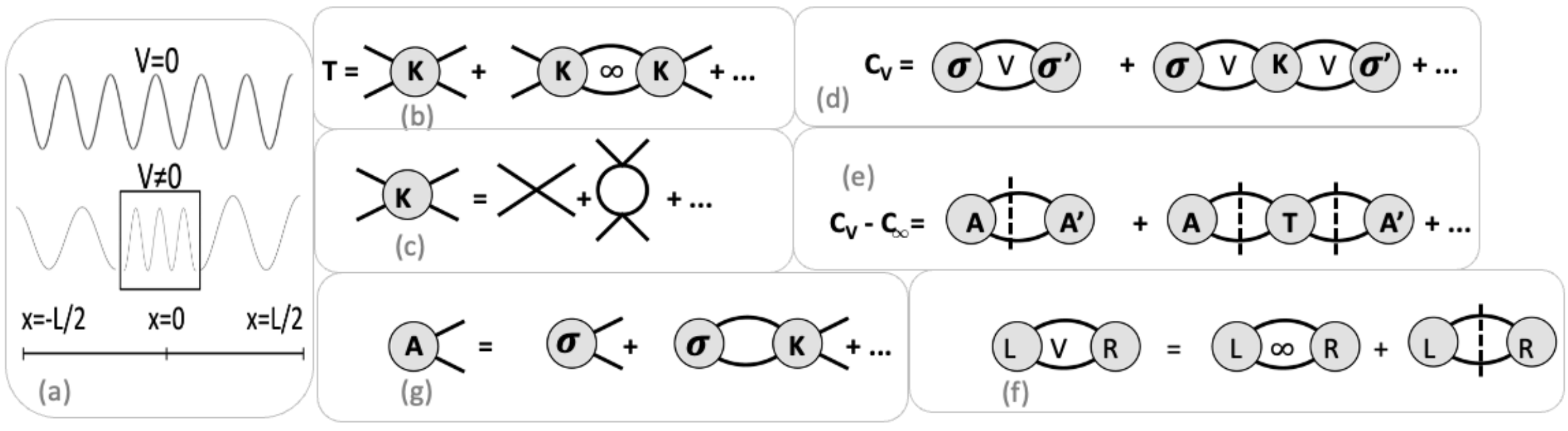}
	\caption{Towards understanding the L\"uscher's relation between finite-volume eigen-energies and infinite volume scattering amplitude $T$ for 1-dimensional quantum mechanics (a) and QFT (b-g) in Section \ref{sec:luscher}. The dashed vertical line denotes the finite-volume correction to the loop function, $\sigma^{(\prime)}$   represent source/sink operators, $A^{(')}$ represent overlaps to dressed source/sink operators, while the remaining notation is described in the text. }
	\label{fig:luscher}
\end{figure}

The relation between eigen-energies $E$ and the scattering amplitude can be most easily derived in one-dimensional quantum mechanics as illustrated in Figure \ref{fig:luscher}(a).  In the absence of potential, the wave function equals $\psi\propto\cos(px)$   with  $p=n\tfrac{2\pi}{L}$ due to the periodic boundary condition. For a non-zero potential of finite range, the wave function  $\psi\propto\cos(px+\delta)$ acquires a phase shift $\delta$ outside the region of potential,  and the momentum is modified to 
\begin{equation}
E\leftrightarrow T(E):\qquad p=n\tfrac{2\pi}{L}-\tfrac{2}{L}\delta \qquad \mathrm{with} \qquad E=p^2/(2m) \qquad T\propto 1/(p\cot\delta - ip)~,
\end{equation}
 in order to still satisfy the periodic boundary conditions $\psi(L/2)=\psi(-L/2)$ and $\psi^\prime(L/2)=\psi^\prime(-L/2)$. The finite volume accompanied by certain boundary conditions, therefore, implies a relation between  $E$ and   $T(E)$ for an interacting system.

  \paragraph{\bf  Derivation of L\" uscher's relation in QFT}\label{sec:luscher-QFT}
  
   Now let's turn to the finite-volume eigen-energies $E_n$   in Quantum Field Theory.  The non-interacting (ni) eigenenergies of two hadrons in a finite box with periodic boundary conditions represent 
   a valuable reference point (which will be denoted by lines in the spectrum plots)
   \begin{equation}
   \label{Eni}
   E^{ni}=\sqrt{\vec p_1^2+m_1^2}+\sqrt{\vec p_2^2+m_2^2}\ , \qquad \vec p_{1,2}=\vec n_{1,2} \tfrac{2\pi}{L}\ ,\quad   \vec n_{1,2}\in Z^3\ , \quad \vec p_1+\vec p_2=\vec P~.
   \end{equation} 
 The energies are modified due to the interactions between two hadrons, and the energy shifts $E-E^{ni}$ depend on the scattering amplitudes. 
  
  In order to derive the relation between eigen-energies $E$ and the scattering amplitude $T(E)$ at the same energy in QFT,   I present essential steps from Ref.  \cite{Kim:2005gf}, simplified to the scattering of two non-identical scalar particles in partial wave $l=0$ and total momentum $P=(E,\vec 0)$. The scattering amplitude $T$ is given by the infinite sum of diagrams presented in Figure \ref{fig:luscher}(b). The lines represent the dressed renormalized scalar propagators, and the kernel $K$ is the sum of two-particle irreducible diagrams in the s-channel illustrated in  Figure \ref{fig:luscher}(c).

  We will relate energy and scattering amplitude via the finite-volume correlation function $C_V$ that is Fourier transformed from  Minkowski time $t_M$ to energy. On one hand, it depends on the eigen-enegies as 
  \begin{equation}
  \label{CV}
  C_V(E)=\int dt_Me^{iEt_M} C_V(t_M)=\int dt_Me^{iEt_M} \sum_n A_n e^{-iE_n t_M}=\sum_n A_n \delta(E-E_n)~,
  \end{equation}
   being singular at each $E_n$. On the other hand, $C_V$ is given by the infinite sum of diagrams with kernel $K$ in Figure \ref{fig:luscher}(d), and our aim is to express $C_V$  in terms of the scattering amplitude. The important contribution to $C_V$ will be represented by the intermediate particles that can go on-shell, propagate to the spatial boundary, and render significant finite-volume effects, while the exponentially suppressed finite-volume corrections $e^{-mL}$ will be neglected \cite{Kim:2005gf}. 
   
   Correlator $C_V$ features the finite-volume loop $I(V)$ in Figure \ref{fig:luscher}(f), which can be decomposed into the sum of the infinite-volume loop $I(\infty)$ and the finite-volume correction $I(V)-I(\infty)$.  The correction arises from the difference between the integral over the loop momenta $\int d^3\vec k$ and the sum over the discrete values in a periodic box $\sum_{\vec k=\vec n 2\pi/L}$ :
  \begin{align}
  I(V)-I(\infty)&=\biggl(\frac{1}{L^3}\!\!\sum_{\vec k}-\int \!\!\frac{d\vec k}{(2\pi)^3}\biggr)\int \!\!dk_0~ iL(|\vec k| )~\frac{i}{k^2-m^2+i\epsilon}~\frac{i}{(P-k)^2-m^2+i\epsilon}~iR(|\vec k| )~, \quad P=(E,\vec 0), \ \ \omega_{\vec k}\equiv\sqrt{m^2+\vec k^2}\nonumber\\
 & =\biggl(\frac{1}{L^3}\!\!\sum_{\vec k}-\int \!\!\frac{d\vec k}{(2\pi)^3}\biggr) \int \!\!dk_0~iL(|\vec k| )~\frac{i}{(k_0-\omega_{\vec k}+i\epsilon)(k_0+\omega_{\vec k}-i\epsilon)}~\frac{i}{(E-k_0-\omega_{\vec k}+i\epsilon)(E-k_0+\omega_{\vec k}-i\epsilon)}~iR(|\vec k| )~.
 \label{L1}
  \end{align}
  Here $L(|\vec k| )$ and $R(|\vec k|)$ represent any of   $K$, $T$ or $\sigma$, and they depend only on the magnitude of the momentum in case of the s-wave scattering.  The integral $\int_{-\infty}^\infty dk_0$ is evaluated by applying the Cauchy theorem. The contour is closed around the lower-half plane and encapsulates the poles at $k_0^{(1)}=\omega_{\vec k}-i \varepsilon$ and $k_0^{(2)}=E+\omega_{\vec k}-i \varepsilon$.   The residue of the integrand at $k_0^{(2)}$ is non-singular in the whole region of interest. This renders only the neglected exponentially suppressed contribution  since $I(V)-I(\infty)\propto (\tfrac{1}{L^3}\!\!\sum_{\vec k}-\int \!\!\tfrac{d\vec k}{(2\pi)^3})f(\vec k)\propto O(e^{-mL})$ for non-singular integrand $f(\vec k)$ according to the Poisson summation formula. The residue of the integrand (\ref{L1})  at the first pole  $k_0^{(1)}$ is proportional to $1/ (E-2\omega_{\vec k}+i\varepsilon)$ and is singular  only when $E=2\omega_{\vec k}$. This occurs when both particles are on-shell, and the loop momentum $|\vec k|$ is equal to the on-shell momentum $p$. The difference  between the finite sum and the integral    for a singular integrand is not exponentially suppressed, and renders an important contribution 
    \begin{align}
   I(V)-I(\infty)&=iL(p)~\underbrace{\biggl\{ \biggl(\frac{1}{L^3}\!\!\sum_{\vec k}-\int \!\!\frac{d\vec k}{(2\pi)^3}\biggr) ~\frac{i }{2\omega_{\vec k}~ E~ (E-2\omega_{\vec k}+i\epsilon)}~\biggr\}}_{-F(E,L)} ~iR(p)=iL(p)~\underbrace{\biggl\{ \biggl(\frac{1}{L^3}\!\!\sum_{\vec k}-\int \!\!\frac{d\vec k}{(2\pi)^3}\biggr) ~\frac{i ~ 4\pi Y_{00}(\hat k)Y_{00}(\hat k) }{2\omega_{\vec k}~ E~ (E-2\omega_{\vec k}+i\epsilon)}~\biggr\}}_{-F(E,L)=-F_{00,00}(E,L)} ~iR(p)~.
   \label{L2}
   \end{align}
  The above equation introduces the kinematical function $F=F_{00,00}$, which will appear in the final L\"uscher's relation, and matches the more general $F_{l_1m_1,l_2m_2}$  defined in Eq. (48) of Ref. \cite{Kim:2005gf}. The same reference also provides a useful relation to evaluate $F$, which is independent of the regularization since the ultra-violet part cancels.   The kernels $L$ and $R$ in  (\ref{L2}) are evaluated at the on-shell momentum $p$ as dictated by the pole  $1/(E-2\omega_{\vec k}+i\epsilon)$. This separation between the on-shell quantities and the kinematical function will allow us to express the correlator $C_V(E)$ in terms of the on-shell scattering amplitude $T$.

  The finite-volume loop, expressed as a sum $I(V)=I(\infty)+[I(V)-I(\infty)]$, is inserted in place of each loop in the finite-volume correlator $C_V(E)$  in Figure  \ref{fig:luscher}(d). A proliferation of terms for $C_V(E)$ is obtained. They can be combined as shown in Figure \ref{fig:luscher}(e), as the reader can check by drawing some examples in this sum.   The first term contains only  $I(\infty)$ and corresponds to the correlation function in infinite volume $C_{\infty}(E)$. The other diagrams can be expressed in terms of  $I(V)-I(\infty)$, the dressed overlaps to operators $A^{(')}$   and the infinite volume scattering amplitude $T$, all defined in Figure \ref{fig:luscher}.   The resulting finite volume correlator 
\begin{equation}
C_V(E)=C_{\infty}(E)+A^\prime[-F+FiTF+..]A=C_{\infty}(E)-A^\prime F\sum_{j=0}^{\infty} (-iTF)^j A = C_{\infty}(E)-A^\prime \frac{1}{T+iF^{-1}}A~
\end{equation}
is now expressed in terms of the on-shell infinite-volume scattering amplitude $T$.  This comes about since the finite-volume correction to the loop in  (\ref{L2}) depends on the on-shell quantities, which follows from the dominant role of on-shell particles in the loop. Poles of $C_V$ appear when the denominator vanishes and where $E=E_n$ according to (\ref{CV}).  This leads to the L\"uscher's relation between $T(E)$ and  $E$ 
\begin{equation}
\label{luscher}
\vec P=0, \ l=0 :\ \  \boxed{T_{l=0}(E)+iF^{-1}(E)=0} \quad \to \quad  \  T_{l=0}^{-1}(E)=-iF(E) ~,
 \end{equation}
 which applies when $E$ is equal to the eigen-energy $E_n$. 
 The kinematical function $F=F_{00,00}$ is given in Eq. (48) of  \cite{Kim:2005gf}. The lattice eigen-energy $E$, therefore, directly renders the scattering amplitude $T(E)$ at the same energy. 
 
\begin{figure}[htb]
	\centering
	\includegraphics[width=0.48\textwidth]{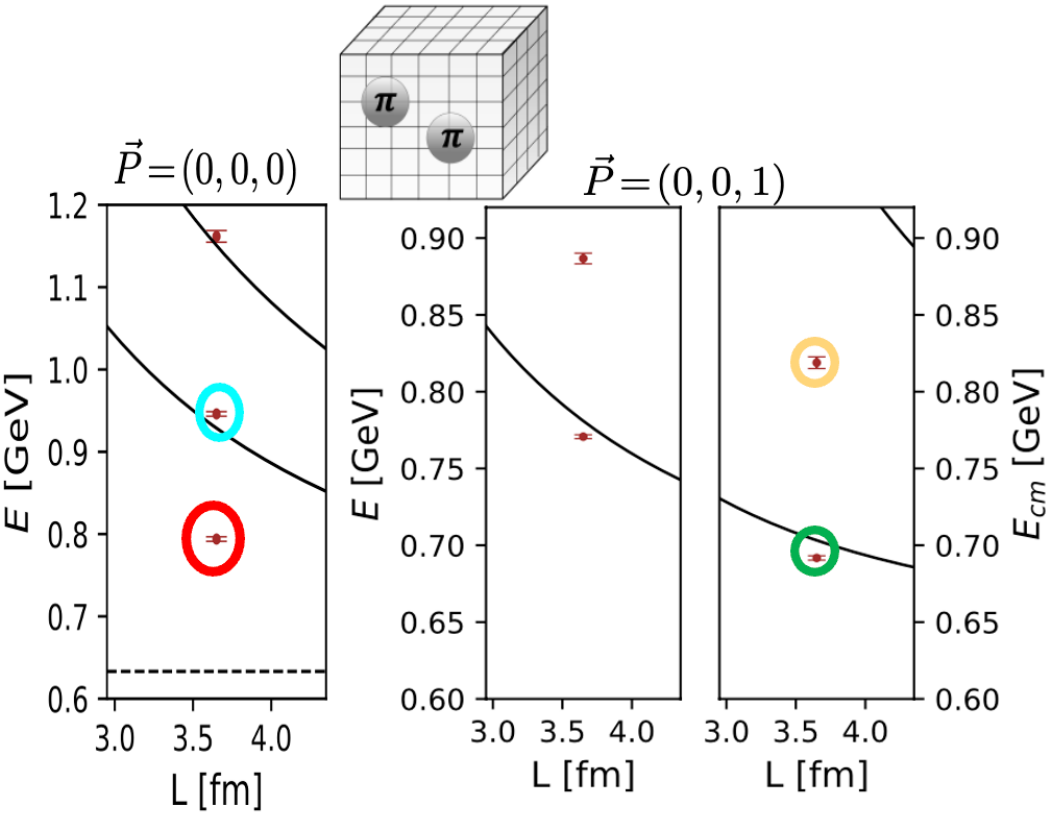}$\quad$
	\includegraphics[width=0.48\textwidth]{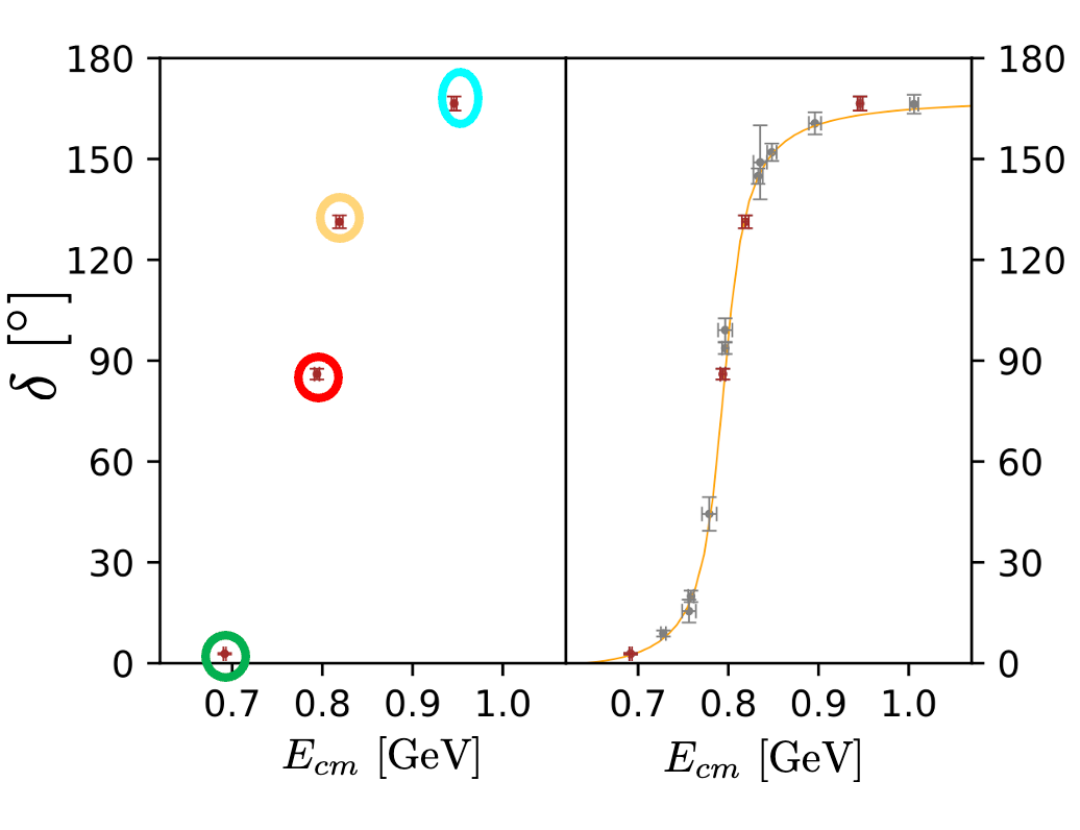}
	\caption{ The $\pi\pi$ scattering in $\rho-$meson channel from a lattice simulation \cite{Alexandrou:2017mpi} at $m_\pi\simeq 320~$MeV: energies and phase shifts. }
	\label{fig:pipi-pedagogical}
\end{figure}

\begin{figure}[htb]
	\centering
	\includegraphics[width=0.99\textwidth]{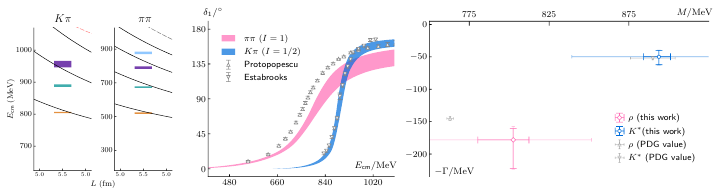}
	\caption{ Resonances $\rho\to\pi\pi$ and $K^*\to K \pi$ from a lattice simulation \cite{Boyle:2024hvv} at physical quark masses: eigenenergies for  $|\vec P|=0$ (left), phase shifts (middle) and the resonance parameters related to the pole position (right).}
	\label{fig:pipi-Kpi-rbc}
\end{figure}

\subsection{Examples} 

\paragraph{\bf Resonances $\rho$ and $K^*$  }

In order to illustrate how resonances are studied in lattice QCD,  the study of $\pi\pi\to\pi\pi$ scattering in the $\rho$-meson channel with $J^P=1^-$ from Ref. \cite{Alexandrou:2017mpi} is presented in Figure \ref{fig:pipi-pedagogical}. The finite-volume energies of an interacting two-pion system on a single volume with $L\simeq 3.6~$fm and $m_\pi\simeq 320~$MeV  are shown for total momenta $|\vec P|=0$ and $1\cdot (\tfrac{2\pi}{L})$. The energies are shifted with respect to the non-interacting energies (represented by lines), which implies non-zero interaction between pions. 
Each energy $E_{cm}$ renders the scattering amplitude and the scattering phase shift at that energy $E_{cm}$, as illustrated by circles of various colors: the underlying relation for ${\vec P}=\vec 0$ is given in Eq. (\ref{luscher})\footnote{This relation applies here although this is scattering with $l=1$. }.  The same lattice study extracted the eigen-energies also for higher total momenta up to $|\vec P|=\sqrt{3} \tfrac{2\pi}{L}$, thereby probing this two-pion system and the corresponding phase shift at further kinematical points $E_{cm}$.   All points are collected in the plot on the right, where the phase shift features a clear resonance shape and is described well with a Breit-Wigner form\footnote{The factor $\tfrac{p}{8\pi E}$ comes from normalization choice for $T$ in Eq. (\ref{T}). }
\begin{equation}
\label{BW}
\tfrac{p}{8\pi E} T(E)=\frac{1}{\cot \delta - i}=\frac{\Gamma(E_{cm}) E_{cm}}{m^2-E_{cm}^2 - i ~\Gamma(E_{cm}) E_{cm}}~,\quad \Gamma(E_{cm})=\frac{g^2}{6\pi}\frac{p^3}{E_{cm}^2}. 
\end{equation}
 The phase shift rises through $\delta=90^\circ$ at the resonance mass $m_\rho^{\scriptscriptstyle{ (m_\pi\simeq 320\mathrm{MeV})}}\!\!=798(7)~$MeV which is close to the $m_\rho \simeq 770~$MeV in Nature. The width is smaller than in Nature due to the smaller phase space for decay $\rho\to \pi\pi$ at heavier-than-physical pions, but the $\rho\pi\pi$ coupling $g=5.7(2)$ that parametrizes the width has the value close to the value $g\simeq 6.0$ in Nature. 
 
 A large number of lattice collaborations have already established the resonances $\rho\to \pi\pi$ and $K^*\to K\pi$   at a variety of $u/d$ and $s$ quark masses, including the physical ones. Figure \ref{fig:pipi-Kpi-rbc} displays recent results employing  RBC/UKQCD ensembles with the physical quark masses   \cite{Boyle:2024hvv}, using Breit-Wigner parametrization (\ref{BW}) and verifying good agreement with experiment for the phase shifts and the pole position $m- \tfrac{i}{2}\Gamma$.

\begin{figure}[htb]
	\centering
	\includegraphics[width=0.8\textwidth]{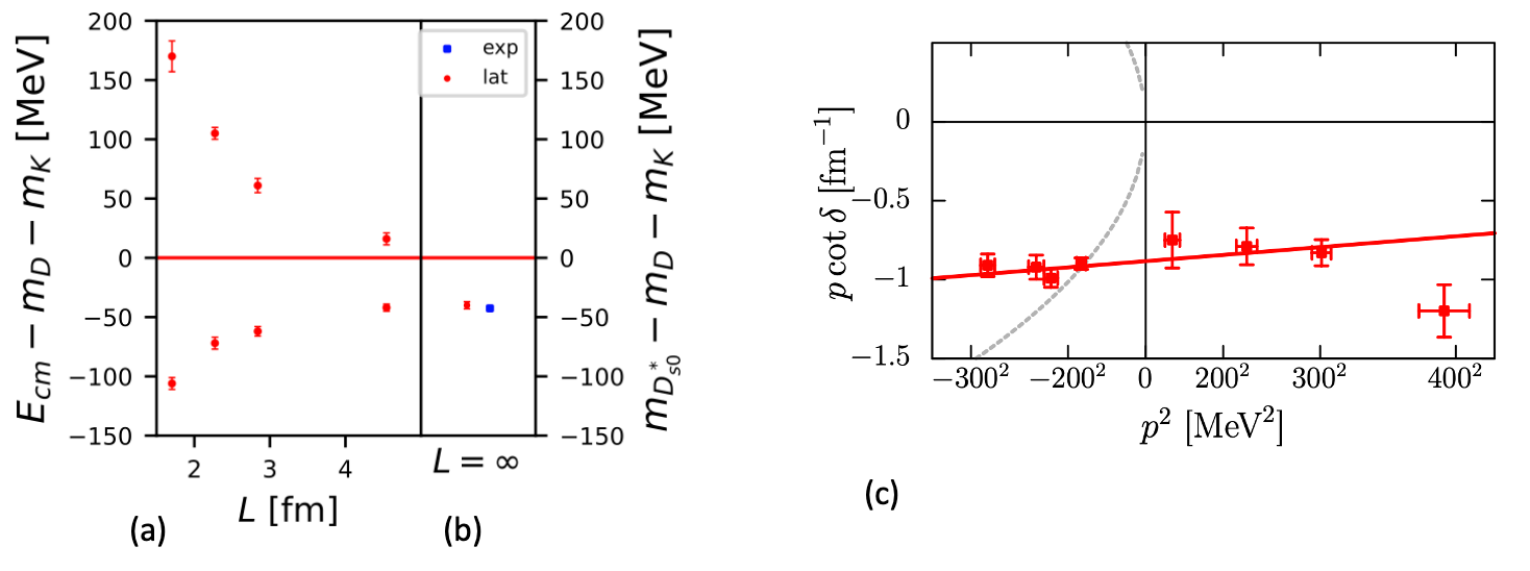}
	\caption{The $DK$ scattering in  with $J^P\!=\!0^+$ and $m_\pi\!=\!290~$MeV  from four  volumes by RQCD \cite{Bali:2017pdv}: (a) finite volume energies at $\vec P=0$, (b) position of the $D_{s0}^*(2317)$ bound state compared to experiment, (c)   $p \cot \delta$ extracted from eight energies via the L\"uscher's relation (\ref{luscher}) is shown in red, while $ip$ is shown in grey.  }
	\label{fig:Ds-RQCD}
\end{figure}

\begin{figure}[htb]
	\centering
	\includegraphics[height=2.3cm]{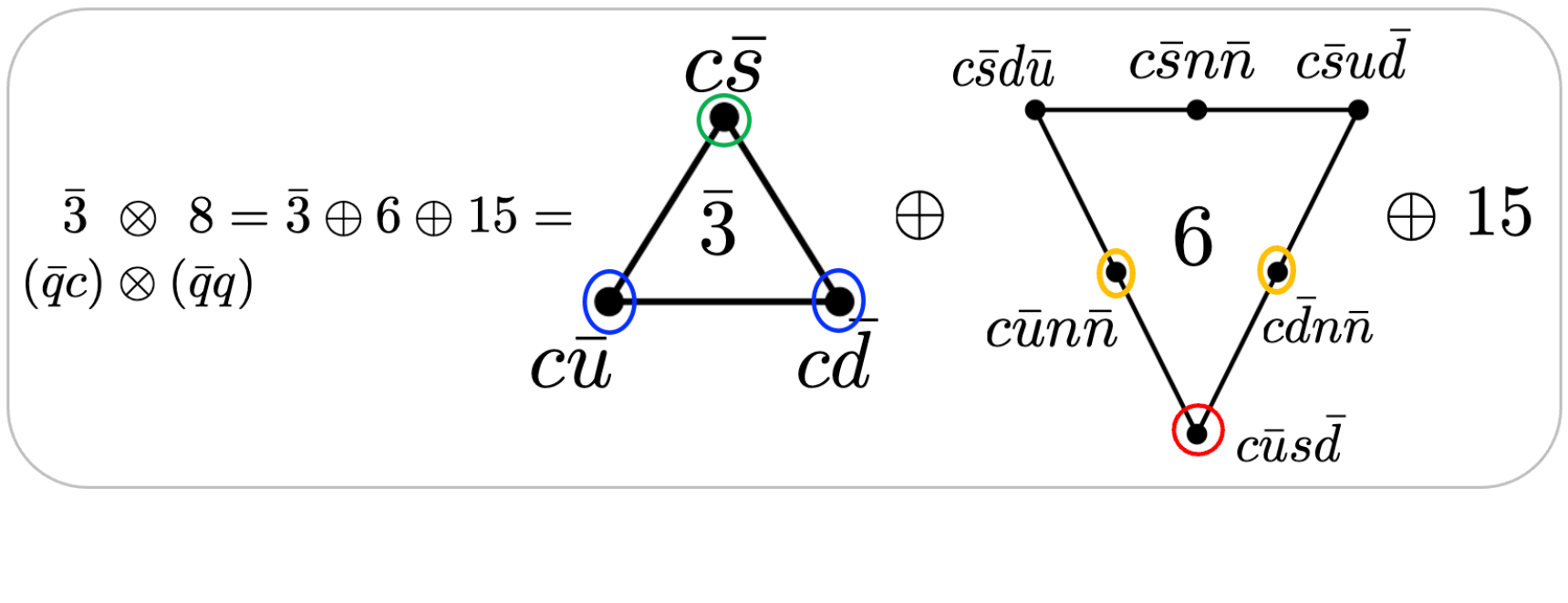}
	$\quad$ \includegraphics[height=2.9cm]{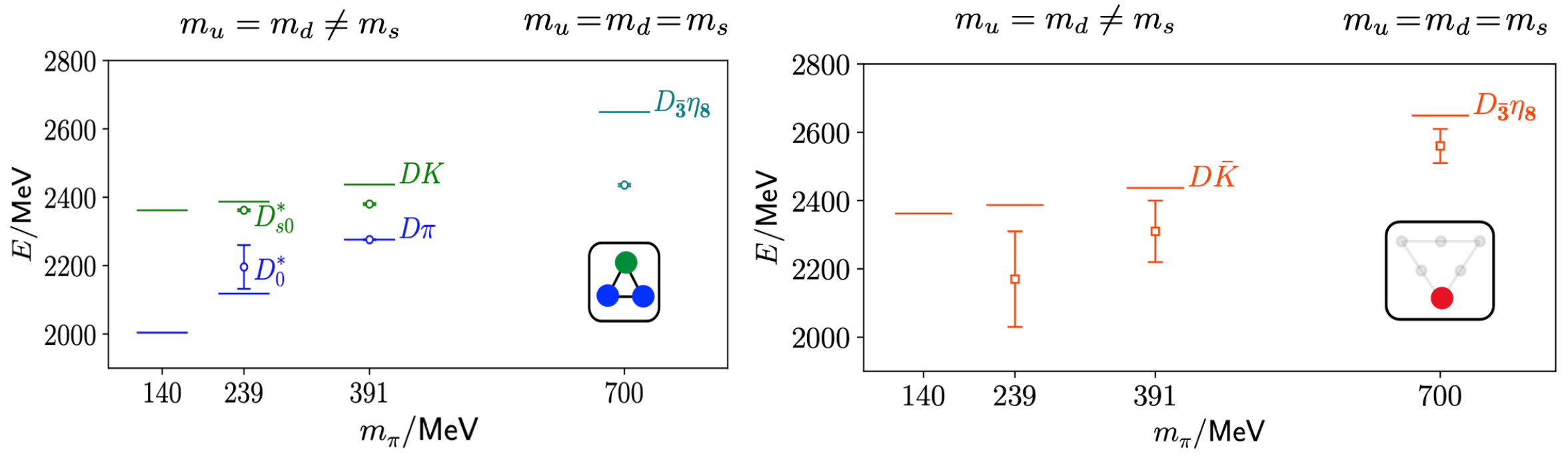}  
	\caption{Charmed scalar mesons: Left: multiplets according to the paradigm that features Fock components  $(c\bar q)$ and $  (c\bar q)(\bar qq)$ with $q=u,d,s$ and $n=u,d$. The states from multiples $\bar 3$ and $6$ can mix when $m_{u/d}\not =m_s$. Right:   Pole locations for three pion masses   (symbols) and relevant thresholds (horizontal lines) 
from lattice studies of Hadron Spectrum collaboration, as summarized in  \cite{Yeo:2024chk}. 	}
	\label{fig:hl}
\end{figure}

\paragraph{\bf Heavy-light  scalar mesons}

The scalar charm-strange meson $D_{s0}^*$ is 
a vanilla example of a bound state. It 
was experimentally discovered about $42~$MeV below $DK$ threshold and is strongly stable in the isospin limit $m_u\!=\!m_d$. The  $DK$ scattering phase shift in partial wave $l=0$ was extracted at eight values of energies,   obtained from four spatial volumes and applying L\"uscher's relation (\ref{luscher}) \cite{Bali:2017pdv}, all shown in Figure \ref{fig:Ds-RQCD}.  As momenta and phase shifts are imaginary below the threshold, it is customary to present the real quantity $p\cot \delta$. The scattering amplitude $T\propto 1/(p \cot\delta - ip)$ (\ref{T}) features a bound state pole below threshold,  where $p\cot \delta$ (red line) intersects with $ip=-|p|$ (dashed gray line) for positive imaginary momenta $p=i|p|$.  The position of this bound state in pane (b) agrees well with the experimental mass of  $D_{s0}^*(2317)$. 

The charmed scalar mesons would form a $SU(3)$ flavor anti-triplet $c\bar q$    with $q=u,d,s$ according to the quark model. However, a new paradigm is supported by a number of studies using effective field theories,  lattice simulations as well as re-analysis of experimental data, for example \cite{Kolomeitsev:2003ac,Du:2017zvv,Albaladejo:2016lbb,Yeo:2024chk}.  According to this paradigm, the spectrum features $c\bar q$ as well as $c\bar q~\bar qq$ Fock components. The latter decomposes to the multiplets $\bar 3\oplus 6\oplus 15$ in the $SU(3)$ flavor limit.   The attractive interactions within the anti-triplet and the sextet suggest the existence of hadrons indicated by circles in Figure \ref{fig:hl}, with two pairs of poles for $I=1/2$ charmed mesons. The spectrum of charmed scalar mesons by the Hadron Spectrum collaboration shows members of the triplet and sextet at three different pion masses, as summarized in \cite{Yeo:2024chk}.  The state $c\bar u s \bar d$ indicated by red carries an exotic flavor quantum number. All these hadrons have been extracted from poles of  $T(E)$ for the scattering of charmed and light mesons with the L\"uscher's formalism. 

\begin{figure}[htb]
	\centering
	\includegraphics[width=0.99\textwidth]{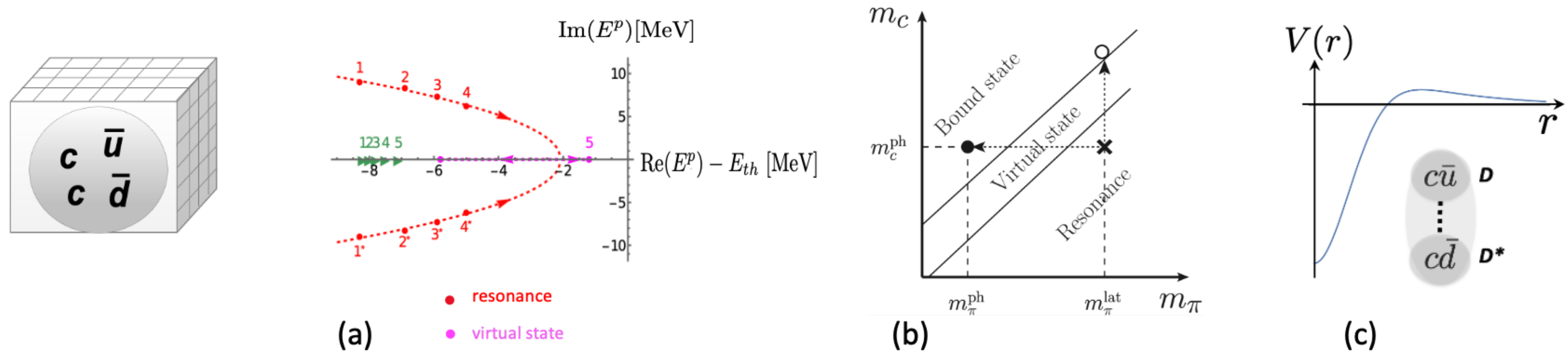}
	\caption{ $T_{cc}=cc\bar u \bar d$ channel with  $J^P=1^+$ and $I=0$ from   \cite{Collins:2024sfi}: (a)   pole trajectory for five charm quark masses at  $m_\pi\simeq 280~$MeV, where only central values are shown; (b)  identity of the pole near $DD^*$ threshold for various charm and light quark masses;  (c) a possible interpretation for kinematics with $m_\pi>m_{D^*}-m_D$. }
	\label{fig:Tcc}
\end{figure}

\paragraph{\bf Doubly heavy tetraquarks   $bc\bar u\bar d$ and $cc\bar u\bar d$}

The doubly charm tetraquark $T_{cc}=cc\bar u\bar d$ with  $J^P=1^+$ and $I=0$ was experimentally discovered less than $1~$MeV below $DD^*$ threshold  by LHCb \cite{LHCb:2021auc}. Several recent lattice simulations have extracted the $DD^*$ scattering amplitude in the kinematics where $D^*$ is stable, and all simulations indeed found the pole near the $DD^*$ threshold. The quark mass dependence of the pole position was investigated in \cite{Collins:2024sfi}, where $DD^*$ scattering amplitude was extracted from finite-volume energies by combining L\"uscher approach and Effective Field Theory to incorporate the effects of the so-called left-hand cut \cite{Du:2023hlu}\footnote{L\"uscher's relation namely does not apply in the region below threshold where pion comes on-shell in u-channel exchange (in kinematics when $m_\pi>m_{D^*}-m_D$).}. The pole transitions between a resonance, virtual state, and bound state when $m_c$ is increased or when  $m_{u/d}$ is decreased, as shown in Figures \ref{fig:Tcc}(a,b). Such a pole trajectory is consistent (but does not uniquely imply)  with $DD^*$ interacting through the potential sketched in Figure \ref{fig:Tcc}(c), which is almost independent of $m_c$ and becomes more attractive with decreasing $m_{u/d}$ \cite{Collins:2024sfi}. This is roughly in line with the expectation from the exchange of light mesons between $D$ and $D^*$.   The current understanding of this interesting state as of 2024 will surely improve with further studies. 

\begin{figure}[htb]
	\centering
	\includegraphics[width=0.99\textwidth]{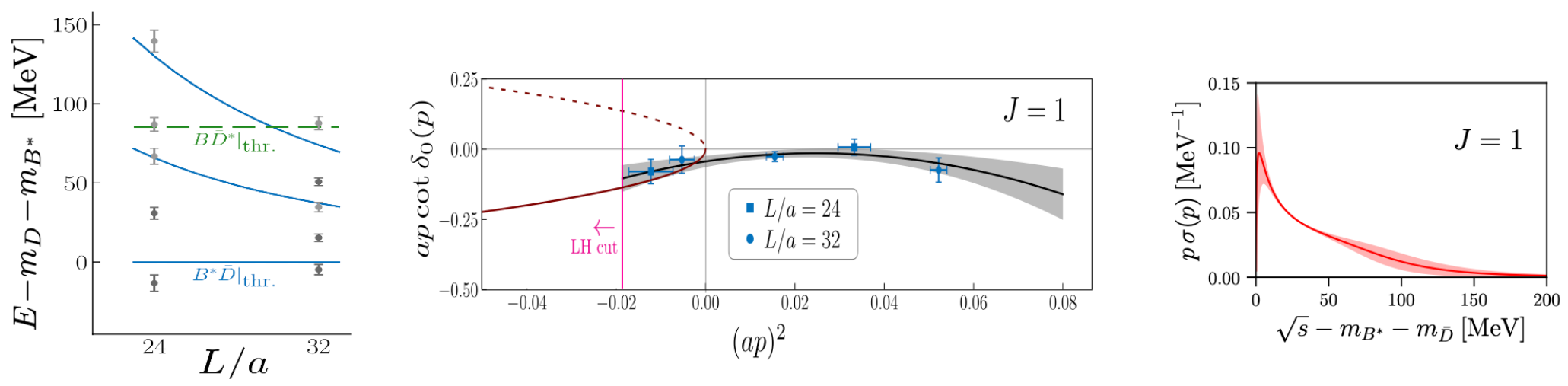}
		\caption{The $T_{bc}=bc\bar u\bar d$ system with $J^P=1^+$ and $I=0$ from simulation  at $m_\pi\simeq 220~$MeV \cite{Alexandrou:2023cqg}. Left:  eigen-energies with $\vec P=\vec 0$; Middle:  $p\cot \delta$ that features a bound state pole where fit crosses the brown line. Right:  $p\sigma$  shows a peak above the threshold due to the bound state pole below the threshold.    }
	\label{fig:Tbc}
\end{figure}

 No other doubly heavy tetraquark has been experimentally discovered, and $T_{bc}=bc\bar u\bar d$ could perhaps be the next discovery in line.  The lattice simulation of this system features a bound state with $J^P=1^+$ and $I=0$, that resides about $2~$MeV below $DB^*$ threshold in lattice simulation \cite{Alexandrou:2023cqg}, as shown in Figure \ref{fig:Tbc}.  
 
\paragraph{\bf Nucleon-nucleon scattering}

\begin{figure}[htb]
	\centering
	\includegraphics[height=4.8cm]{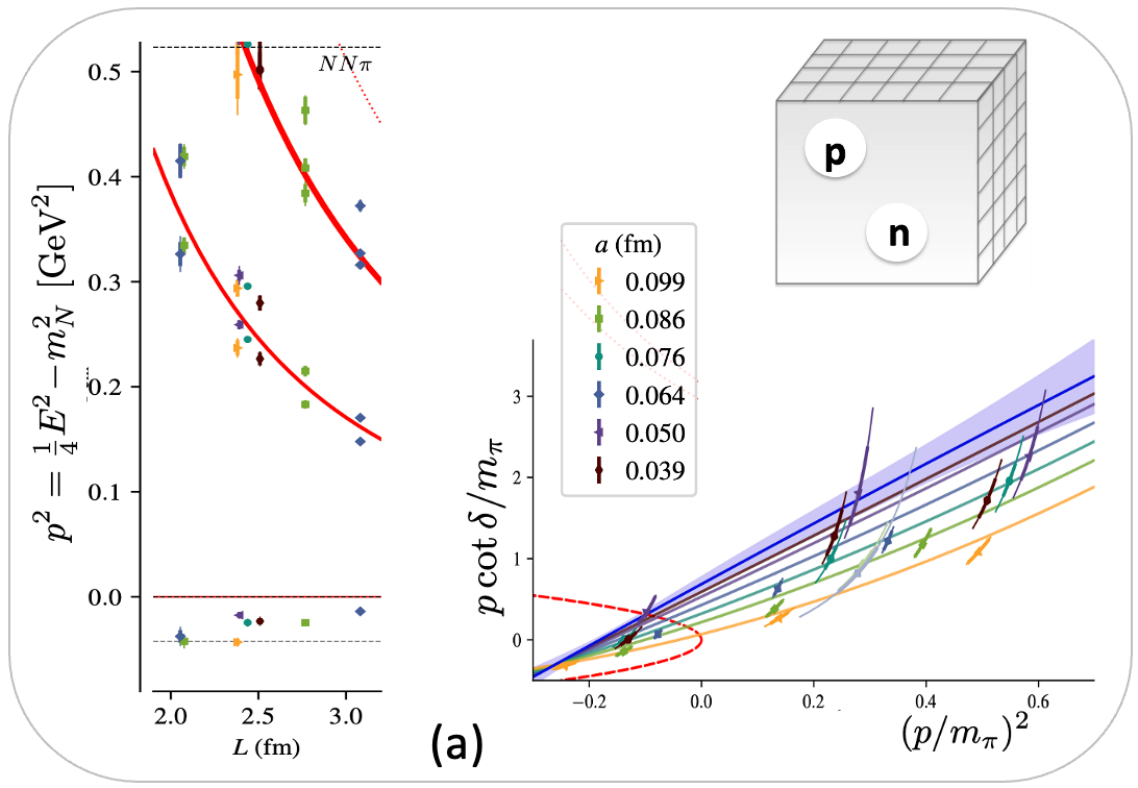}  \includegraphics[height=4.8cm]{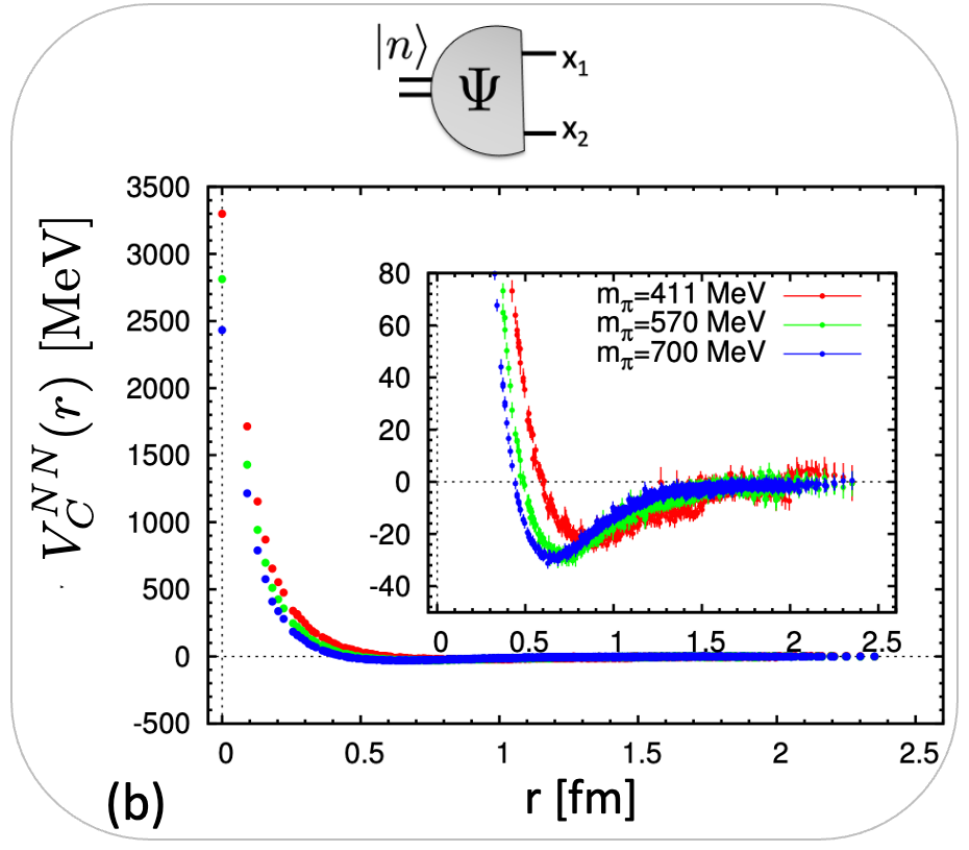}\includegraphics[height=4.8cm]{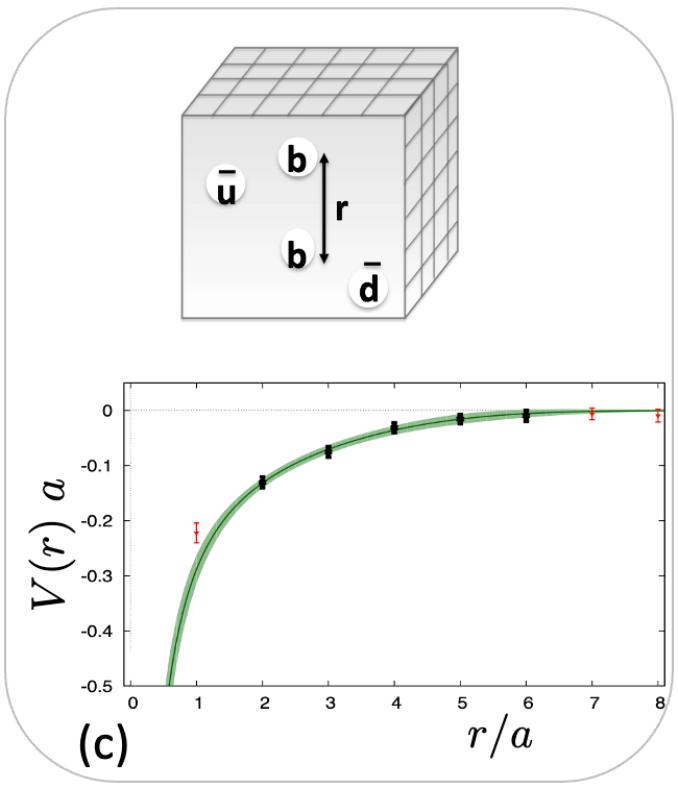}
	\caption{ (a) Deuteron channel $^{2S+1}L_J=^3S_1$ from the simulation with  $m_u=m_d=m_s$ and $m_\pi\simeq 420~$MeV: eigen-energies at total momentum zero, $p \cot\delta $ and its dependence on the lattice spacing $a$  \cite{Green:2022rjj}. (b) Deuteron channel from HAL QCD approach that determines the central part of the local two-nucleon potential $V$ from the Bethe-Salpeter wave function $\Psi$ \cite{Ishii:2013ira,Aoki:2023qih}.  (c) 
     Potential between $B$ and $B^*$ for the system $bb\bar u\bar d$ with $J^P=1^+$ obtained with static $b$ quarks and $a=0.079~$fm \cite{Bicudo:2012qt}. }
	\label{fig:various}
\end{figure}

The scattering of nucleons is particularly challenging due to the signal-to-noise problem. In order to extract the amplitude for the scattering of proton and neutron in the deuteron channel, the eigen-energies of this system with $J^P=1^+$ have been determined at zero total momentum (shown in Figure \ref{fig:various}a) and also at other total momenta. The generalization of the L\"uscher's relation renders $ p\cot \delta$ for partial wave $l=0$  in Figure \ref{fig:various}a, in the approximation where  $l=2$ is neglected.  The scattering amplitude (\ref{T}) has a virtual state pole where $ p\cot \delta=ip$. This is realized for $p=-i|p|$ where the $ p\cot \delta$ line intersects with the red dashed line representing $ip=+|p|$. This simulation at $m_\pi\simeq 420~$MeV confirms the attraction between proton and neutron, which is, however, not large enough to render a deutron bound state, but is responsible for a near-threshold virtual state.  
Different colors correspond to simulations at different lattice spacings $a$, which emphasizes significant discretization effects for this channel.

\begin{figure}[htb]
	\centering
	\includegraphics[height=4.5cm]{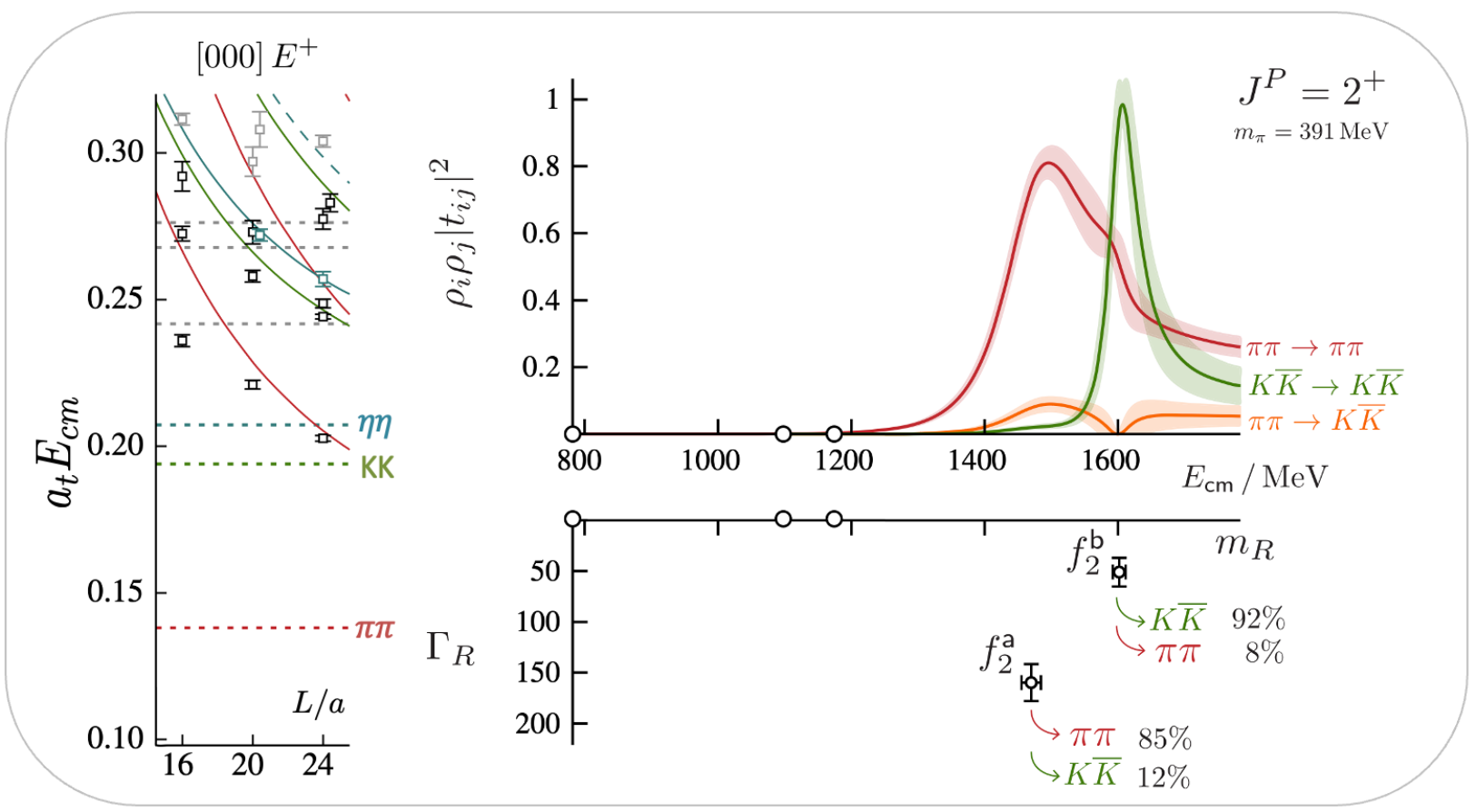}  $\ \ $ \includegraphics[height=4.5cm]{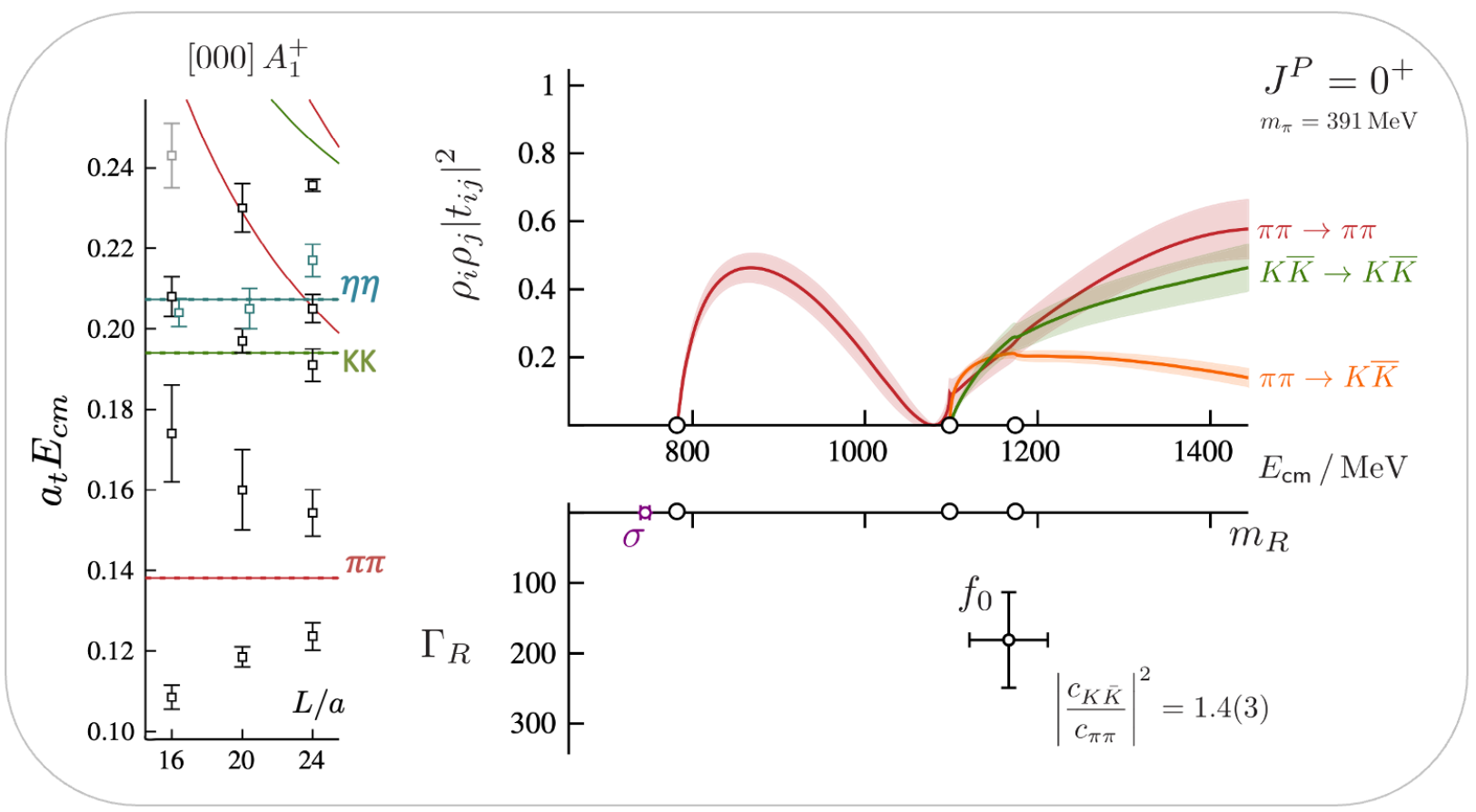}
	\caption{The coupled $\pi\pi-K\bar K-\eta\eta$ scattering at $m_\pi\simeq 390~$MeV in partial waves $l=0$ (left) and $l=2$ (right) from HadSpec collaboration \cite{Briceno:2017qmb}: eigen-energies for three volumes at $\vec P=\vec 0$, quantity related to scattering rates  $\rho\rho |t|^2 $ and pole locations of  $\sigma$, $f_0$ and $f_2$. Red, green and blue lines represent non-interacting energies of  $\pi\pi$, $K\bar K$ and $\eta\eta$, respectively. Here $\rho_i\equiv 2p_i/E_{cm}$ and $t_{ij}= T_{ij}/16\pi$ (\ref{T}).}
	\label{fig:coupled}
\end{figure}

\section{Hadrons from coupled-channel scattering}\label{sec:coupled}

Most resonances lie above several thresholds and decay to several final states via the strong interaction. Light-meson resonances $f_0$ and $f_2$ that feature in the scattering of three coupled channels $\pi\pi-K\bar K-\eta\eta$ are shown in Figure \ref{fig:coupled} \cite{Briceno:2017qmb}. The information on the scattering is again obtained from the finite volume eigen-energies, where the interacting two-meson states $\pi\pi$, $K\bar K$, and $\eta\eta$ feature in the spectra, shown on the left. The scattering is described with $3\times 3$ scattering matrix, where each element $T_{ij}(E)=T_{ji}(E)$ depends on the energy. The L\"uscher's equation for one-channel (\ref{luscher}) generalizes  for the case of three coupled channels with spinless particles in partial wave $l=0$ to 
\begin{equation}
\label{coupled}
\det\Biggl[~ \begin{pmatrix} T_{aa}(E) & T_{ab}(E)  & T_{ac}(E)\\  
                                T_{ab}(E) & T_{bb}(E)  & T_{bc}(E)\\  
                                T_{ac}(E) & T_{bc}(E)  & T_{cc}(E) \end{pmatrix} + 
                                i \begin{pmatrix}  F^{-1}_{a}(E) &  0 & 0 \\  
                                 0 & F^{-1}_{b}(E)  &  0 \\  
                              0 & 0   &  F^{-1}_{c}(E) \end{pmatrix}~\Biggr] = 0 \ ,\qquad  \begin{matrix}a=\pi\pi\\ b=K\bar K\\ c=\eta\eta \end{matrix}
\end{equation}
 and it relates $E$ and $T_{ij}(E)$ when $E$ is lattice eigen-energy. 
Here, $F_i(E)$ are the same kinematical functions as for one-channel scattering \cite{Kim:2005gf}, and they depend on the masses of the scattering particles in channel $i$.  For a given lattice eigen-energy $E$, the determinant equation is one equation for six unknown $T_{ij}(E)$. It is, therefore, customary to parametrize the energy dependence of all $T_{ij}^{model}(E,\vec \kappa)$ in terms of a few unknown parameters $\vec \kappa$. These parameters are fitted so that lattice eigen-energies are best reproduced with the prediction of the L\"uscher's equation for $T_{ij}^{model}(E,\vec \kappa)$ (\ref{coupled}). The resulting scattering matrix in Figure \ref{fig:coupled}  is constrained by the spectra with total momentum zero,  as well as other total momenta that are not shown.  Two channels $\pi\pi$ and $K\bar K$ are found to be significantly coupled for $l=0$ and almost uncoupled for $l=2$, while coupling to $\eta\eta$ is small in both cases. The poles are related to scalar and tensor resonances $f_0$ and $f_2$, while $\sigma$ pole is below $\pi\pi$ threshold at this pion mass.

\section{Hadrons from static potentials}\label{sec:static-potentials}

Hadronic systems that contain two heavy quarks, i.e. $bb$ or $\bar bb$, and additional light degrees of freedom  (gluons $G$ and/or light quarks $q=u,d,s$), can be addressed via the Born-Oppenheimer approximation since the velocities of heavy quarks are much smaller than those of the light degrees of freedom.  Let's consider a system   $bb\bar u\bar d$  with $J^P=1^+$ in  Figure \ref{fig:various}(c), where 
the eigen-energies of a system with a static pair of $b$ quarks at a fixed distance $r$ were determined in \cite{Bicudo:2012qt}. These are called the static energies and are obtained from the correlation functions as described in Section \ref{sec:En} with the sole exception that spatial positions of heavy quarks are fixed in the operator and the propagator. The static eigen-energies $E(r)$  provide  the potential $V(r)=E(r)-m_B-m_{B^*}$ between $B$ and $B^*$  since the kinetic energies of the static $B^{(*)}$ mesons are zero. The extracted potential in Figure  \ref{fig:various}(c) shows significant attraction at explored distances $r/a=1-8$.   The potential needs to be extrapolated towards $r<a$ and large $r$, which can be done reliably if its analytic form is known. The $BB^*$ scattering amplitudes 
and possible existence of bound states or resonances are then explored via the Schr\"odinger equation $[-\nabla^2/2m_r+V(r)]\psi=W \psi$  like in non-relativistic quantum mechanics. This renders the $T_{bb}$ bound state with the binding energy  $W=-38(18)~$MeV with respect to   $BB^*$  threshold \cite{Bicudo:2012qt}. The formalism for a great variety of exotic channels is worked out in Ref. \cite{Berwein:2024ztx}.
  
\section{Hadrons and potentials from  HAL QCD approach}\label{sec:halqcd}

Eigen-energies represent the main quantity extracted from lattice simulations 
 in the previous sections. On the other hand, the HAL QCD approach is based on extracting the Bethe-Salpeter wave function $\psi$ of an eigenstate directly from lattice QCD.  The wave function of two scalars within an eigenstate $|n\rangle$    is  defined as $\Psi(x_1,x_2)=\langle 0|T[\phi(x_1)\phi(x_2)]|n,E\rangle$  and  illustrated in Figure \ref{fig:various}(b)  \cite{Nakanishi:1969ph}.  
  The  HAL QCD collaboration   investigates  two-hadron systems, where the equal-time Bethe-Salpeter wave function is analogously defined as 
\begin{equation}
\label{psi}
\psi^{H_1H_2}_{E_n}(\vec r) ~e^{-E_nt}\propto  \langle 0| ~H_1(\vec r,t)H_2(\vec 0,t) ~|H_1H_2,E\rangle~. 
\end{equation}
Here $|H_1H_2,E\rangle $ is a two-hadron eigenstate with energy $E$, $H_{1,2}(\vec r,t)$ are single-hadron operators and spins of hadrons are omitted for simplicity. The wave-function $ \psi^{H_1H_2}(\vec r)$ of the ground state can be determined from the large-time behavior of the correlator  \cite{Aoki:2023qih}\footnote{In practice, HAL QCD uses a slightly different method to extract $\psi$ from $F$ - the so-called so-called time-dependent method.}
\begin{equation}
F^{H_1H_2}(\vec r,t)\equiv \textstyle{\sum_{\vec x}}~ \langle 0|H_1(\vec x+\vec r,t+t_0)H_2(\vec x,t+t_0)~J^\dagger_{H_1H_2}(t_0)|0\rangle \stackrel{t\to \infty}{=} A ~\psi^{H_1H_2}(\vec r) ~e^{-Et}~,
\end{equation}
where $J^\dagger_{H_1H_2}(t_0)$ is a source operator that creates two-hadron system at time $t_0$.

The HAL QCD approach relates the wave function $\psi$ to the potential between two hadrons. Within the relativistic theory, this approach renders the so-called non-local potential $U(\vec r,\vec r^\prime)$. Its precise definition goes beyond this introductory text, but it is discussed in numerous   HAL QCD papers, for example, in review \cite{Aoki:2023qih}. The local and central part  of the potential $U(\vec r,\vec r^\prime)=V_C(\vec r)\delta^{(3)}(\vec r-\vec r^\prime)+... $ in the case of a nucleon-nucleon system in the deuteron channel is presented in Figure \ref{fig:various}(b).  Here the standard nucleon operators $H_{1,2}(x) \simeq \epsilon_{abc} [u^a(x)C\gamma_5 d^b(x)] q^c(x)$ are employed   \cite{Ishii:2013ira,Aoki:2023qih}.  The potential is repulsive at short distances, attractive at medium and long distances, and does not render a deuteron bound state at $m_\pi\geq 411~$MeV. This agrees with the absence of the deuteron bound state at $m_\pi \simeq 420~$MeV in the study employing the  L\"uscher's method shown in Figure \ref{fig:various}(a) \cite{Green:2022rjj}. The deuteron bound state is expected to emerge once the pion mass is decreased towards the physical pion mass, in which case the lattice results are currently too noisy to render a reliable conclusion.

\section{Conclusions}
Experiments have provided great discoveries of new conventional hadrons as well as around thirty exotic hadrons. I have presented the theoretical challenge to understand the spectroscopic properties of 
various hadron sectors from ab initio lattice QCD. This approach yields masses of hadrons that are strongly stable, as well as hadrons that are slightly below the strong decay threshold or decay strongly via one decay channel. The theoretical challenge increases with the number of open decay channels. It seems impossible to address the high-lying states like $Z_c(4430)$ with current lattice methods, while many interesting physics conclusions are already available for certain lower-lying states.

\vspace{0.5cm}

{\bf Acknowledgments: } I would like to gratefully thank Christine Davies, Takumi Doi, Sara Collins, Felix Erben, Jeremy Green, Feng-Kun Guo, Tetsuo Hatsuda, Jamie Hudspith, Nelson Pitanga Lachini,   Luka Leskovec, Yan Lyu, William Parrott, Stefan Meinel and  M. Padmanath. Support by Slovenian Research Agency ARIS for funding programme P1-0035 is acknowledged. 

\providecommand{\href}[2]{#2}\begingroup\raggedright\endgroup


\begin{thebibliography}{10}

\bibitem{Brambilla:2019esw}
N.~Brambilla, S.~Eidelman, C.~Hanhart, A.~Nefediev, C.-P. Shen, C.~E. Thomas,
  A.~Vairo, and C.-Z. Yuan, ``{The $XYZ$ states: experimental and theoretical
  status and perspectives},''
  \href{https://dx.doi.org/10.1016/j.physrep.2020.05.001}{{\em Phys. Rept.}
  {\bfseries 873} (2020) 1--154},
  \href{https://arxiv.org/abs/1907.07583}{{\ttfamily arXiv:1907.07583
  [hep-ex]}}.

\bibitem{Bicudo:2022cqi}
P.~Bicudo, ``{Tetraquarks and pentaquarks in lattice QCD with light and heavy
  quarks},'' \href{https://dx.doi.org/10.1016/j.physrep.2023.10.001}{{\em Phys.
  Rept.} {\bfseries 1039} (2023) 1--49},
  \href{https://arxiv.org/abs/2212.07793}{{\ttfamily arXiv:2212.07793
  [hep-lat]}}.

\bibitem{Bulava:2022ovd}
J.~Bulava {\em et~al.}, ``{Hadron Spectroscopy with Lattice QCD},'' in {\em
  {Snowmass 2021}}.
\newblock 3, 2022.
\newblock \href{https://arxiv.org/abs/2203.03230}{{\ttfamily arXiv:2203.03230
  [hep-lat]}}.

\bibitem{Briceno:2017max}
R.~A. Briceno, J.~J. Dudek, and R.~D. Young, ``{Scattering processes and
  resonances from lattice QCD},''
  \href{https://dx.doi.org/10.1103/RevModPhys.90.025001}{{\em Rev. Mod. Phys.}
  {\bfseries 90} no.~2, (2018) 025001},
  \href{https://arxiv.org/abs/1706.06223}{{\ttfamily arXiv:1706.06223
  [hep-lat]}}.

\bibitem{Hanlon:2024fjd}
A.~D. Hanlon, ``{Hadron spectroscopy and few-body dynamics from Lattice QCD},''
  \href{https://dx.doi.org/10.22323/1.453.0106}{{\em PoS} {\bfseries
  LATTICE2023} (2024) 106}, \href{https://arxiv.org/abs/2402.05185}{{\ttfamily
  arXiv:2402.05185 [hep-lat]}}.

\bibitem{pdg2024}
{\bfseries Particle Data Group} Collaboration, S.~Navas {\em et~al.}, ``{Review
  of particle physics},''
  \href{https://dx.doi.org/10.1103/PhysRevD.110.030001}{{\em Phys. Rev. D}
  {\bfseries 110} no.~3, (2024) 030001}.

\bibitem{Gattringer:2010zz}
C.~Gattringer and C.~B. Lang,
  \href{https://dx.doi.org/10.1007/978-3-642-01850-3}{{\em {Quantum
  chromodynamics on the lattice}}}, vol.~788.
\newblock Springer, Berlin, 2010.

\bibitem{Luscher:1990ck}
M.~Luscher and U.~Wolff, ``{How to Calculate the Elastic Scattering Matrix in
  Two-dimensional Quantum Field Theories by Numerical Simulation},''
  \href{https://dx.doi.org/10.1016/0550-3213(90)90540-T}{{\em Nucl. Phys. B}
  {\bfseries 339} (1990) 222--252}.

\bibitem{BMW:2008jgk}
{\bfseries BMW} Collaboration, S.~Durr {\em et~al.}, ``{Ab-Initio Determination
  of Light Hadron Masses},''
  \href{https://dx.doi.org/10.1126/science.1163233}{{\em Science} {\bfseries
  322} (2008) 1224--1227}, \href{https://arxiv.org/abs/0906.3599}{{\ttfamily
  arXiv:0906.3599 [hep-lat]}}.

\bibitem{Dowdall:2012ab}
R.~J. Dowdall, C.~T.~H. Davies, T.~C. Hammant, and R.~R. Horgan, ``{Precise
  heavy-light meson masses and hyperfine splittings from lattice QCD including
  charm quarks in the sea},''
  \href{https://dx.doi.org/10.1103/PhysRevD.86.094510}{{\em Phys. Rev. D}
  {\bfseries 86} (2012) 094510},
  \href{https://arxiv.org/abs/1207.5149}{{\ttfamily arXiv:1207.5149
  [hep-lat]}}.

\bibitem{Colquhoun:2024jzh}
B.~Colquhoun, A.~Francis, R.~J. Hudspith, R.~Lewis, K.~Maltman, and W.~G.
  Parrott, ``{Improved analysis of strong-interaction-stable doubly bottom
  tetraquarks on the lattice},''
  \href{https://dx.doi.org/10.1103/PhysRevD.110.094503}{{\em Phys. Rev. D}
  {\bfseries 110} no.~9, (2024) 094503},
  \href{https://arxiv.org/abs/2407.08816}{{\ttfamily arXiv:2407.08816
  [hep-lat]}}.

\bibitem{Kim:2005gf}
C.~h. Kim, C.~T. Sachrajda, and S.~R. Sharpe, ``{Finite-volume effects for
  two-hadron states in moving frames},''
  \href{https://dx.doi.org/10.1016/j.nuclphysb.2005.08.029}{{\em Nucl. Phys. B}
  {\bfseries 727} (2005) 218--243},
  \href{https://arxiv.org/abs/hep-lat/0507006}{{\ttfamily
  arXiv:hep-lat/0507006}}.

\bibitem{Luscher:1990ux}
M.~Luscher, ``{Two particle states on a torus and their relation to the
  scattering matrix},''
  \href{https://dx.doi.org/10.1016/0550-3213(91)90366-6}{{\em Nucl. Phys. B}
  {\bfseries 354} (1991) 531--578}.

\bibitem{Briceno:2014oea}
R.~A. Briceno, ``{Two-particle multichannel systems in a finite volume with
  arbitrary spin},'' \href{https://dx.doi.org/10.1103/PhysRevD.89.074507}{{\em
  Phys. Rev. D} {\bfseries 89} no.~7, (2014) 074507},
  \href{https://arxiv.org/abs/1401.3312}{{\ttfamily arXiv:1401.3312
  [hep-lat]}}.

\bibitem{Alexandrou:2017mpi}
C.~Alexandrou, L.~Leskovec, S.~Meinel, J.~Negele, S.~Paul, M.~Petschlies,
  A.~Pochinsky, G.~Rendon, and S.~Syritsyn, ``{$P$-wave $\pi\pi$ scattering and
  the $\rho$ resonance from lattice QCD},''
  \href{https://dx.doi.org/10.1103/PhysRevD.96.034525}{{\em Phys. Rev. D}
  {\bfseries 96} no.~3, (2017) 034525},
  \href{https://arxiv.org/abs/1704.05439}{{\ttfamily arXiv:1704.05439
  [hep-lat]}}.

\bibitem{Boyle:2024hvv}
P.~Boyle, F.~Erben, V.~G\"ulpers, M.~T. Hansen, F.~Joswig, M.~Marshall, N.~P.
  Lachini, and A.~Portelli, ``{Light and Strange Vector Resonances from Lattice
  QCD at Physical Quark Masses},''
  \href{https://dx.doi.org/10.1103/PhysRevLett.134.111901}{{\em Phys. Rev.
  Lett.} {\bfseries 134} no.~11, (2025) 111901},
  \href{https://arxiv.org/abs/2406.19194}{{\ttfamily arXiv:2406.19194
  [hep-lat]}}.

\bibitem{Bali:2017pdv}
G.~S. Bali, S.~Collins, A.~Cox, and A.~Sch\"afer, ``{Masses and decay constants
  of the $D_{s0}^*(2317)$ and $D_{s1}(2460)$ from $N_f=2$ lattice QCD close to
  the physical point},''
  \href{https://dx.doi.org/10.1103/PhysRevD.96.074501}{{\em Phys. Rev. D}
  {\bfseries 96} no.~7, (2017) 074501},
  \href{https://arxiv.org/abs/1706.01247}{{\ttfamily arXiv:1706.01247
  [hep-lat]}}.

\bibitem{Yeo:2024chk}
{\bfseries Hadron Spectrum} Collaboration, J.~D.~E. Yeo, C.~E. Thomas, and
  D.~J. Wilson, ``{DK/D\ensuremath{\pi} scattering and an exotic virtual bound
  state at the SU(3) flavour symmetric point from lattice QCD},''
  \href{https://dx.doi.org/10.1007/JHEP07(2024)012}{{\em JHEP} {\bfseries 07}
  (2024) 012}, \href{https://arxiv.org/abs/2403.10498}{{\ttfamily
  arXiv:2403.10498 [hep-lat]}}.

\bibitem{Kolomeitsev:2003ac}
E.~E. Kolomeitsev and M.~F.~M. Lutz, ``{On Heavy light meson resonances and
  chiral symmetry},''
  \href{https://dx.doi.org/10.1016/j.physletb.2003.10.118}{{\em Phys. Lett. B}
  {\bfseries 582} (2004) 39--48},
  \href{https://arxiv.org/abs/hep-ph/0307133}{{\ttfamily
  arXiv:hep-ph/0307133}}.

\bibitem{Du:2017zvv}
M.-L. Du, M.~Albaladejo, P.~Fern\'andez-Soler, F.-K. Guo, C.~Hanhart, U.-G.
  Mei\ss{}ner, J.~Nieves, and D.-L. Yao, ``{Towards a new paradigm for
  heavy-light meson spectroscopy},''
  \href{https://dx.doi.org/10.1103/PhysRevD.98.094018}{{\em Phys. Rev. D}
  {\bfseries 98} no.~9, (2018) 094018},
  \href{https://arxiv.org/abs/1712.07957}{{\ttfamily arXiv:1712.07957
  [hep-ph]}}.

\bibitem{Albaladejo:2016lbb}
M.~Albaladejo, P.~Fernandez-Soler, F.-K. Guo, and J.~Nieves, ``{Two-pole
  structure of the $D^\ast_0(2400)$},''
  \href{https://dx.doi.org/10.1016/j.physletb.2017.02.036}{{\em Phys. Lett. B}
  {\bfseries 767} (2017) 465--469},
  \href{https://arxiv.org/abs/1610.06727}{{\ttfamily arXiv:1610.06727
  [hep-ph]}}.

\bibitem{Collins:2024sfi}
S.~Collins, A.~Nefediev, M.~Padmanath, and S.~Prelovsek, ``{Toward the quark
  mass dependence of Tcc+ from lattice QCD},''
  \href{https://dx.doi.org/10.1103/PhysRevD.109.094509}{{\em Phys. Rev. D}
  {\bfseries 109} no.~9, (2024) 094509},
  \href{https://arxiv.org/abs/2402.14715}{{\ttfamily arXiv:2402.14715
  [hep-lat]}}.

\bibitem{LHCb:2021auc}
{\bfseries LHCb} Collaboration, R.~Aaij {\em et~al.}, ``{Study of the doubly
  charmed tetraquark $T_{cc}^{+}$},''
  \href{https://dx.doi.org/10.1038/s41467-022-30206-w}{{\em Nature Commun.}
  {\bfseries 13} no.~1, (2022) 3351},
  \href{https://arxiv.org/abs/2109.01056}{{\ttfamily arXiv:2109.01056
  [hep-ex]}}.

\bibitem{Du:2023hlu}
M.-L. Du, A.~Filin, V.~Baru, X.-K. Dong, E.~Epelbaum, F.-K. Guo, C.~Hanhart,
  A.~Nefediev, J.~Nieves, and Q.~Wang, ``{Role of Left-Hand Cut Contributions
  on Pole Extractions from Lattice Data: Case Study for Tcc(3875)+},''
  \href{https://dx.doi.org/10.1103/PhysRevLett.131.131903}{{\em Phys. Rev.
  Lett.} {\bfseries 131} no.~13, (2023) 131903},
  \href{https://arxiv.org/abs/2303.09441}{{\ttfamily arXiv:2303.09441
  [hep-ph]}}.

\bibitem{Alexandrou:2023cqg}
C.~Alexandrou, J.~Finkenrath, T.~Leontiou, S.~Meinel, M.~Pflaumer, and
  M.~Wagner, ``{Shallow Bound States and Hints for Broad Resonances with Quark
  Content b\textasciimacron{}c\textasciimacron{}ud in B-D\textasciimacron{} and
  B*-D\textasciimacron{} Scattering from Lattice QCD},''
  \href{https://dx.doi.org/10.1103/PhysRevLett.132.151902}{{\em Phys. Rev.
  Lett.} {\bfseries 132} no.~15, (2024) 151902},
  \href{https://arxiv.org/abs/2312.02925}{{\ttfamily arXiv:2312.02925
  [hep-lat]}}.

\bibitem{Green:2022rjj}
{\bfseries Baryon Scattering (BaSc)} Collaboration, J.~R. Green, A.~D. Hanlon,
  P.~M. Junnarkar, and H.~Wittig, ``{Nucleon-nucleon scattering from
  distillation},'' \href{https://dx.doi.org/10.22323/1.430.0200}{{\em PoS}
  {\bfseries LATTICE2022} (2023) 200},
  \href{https://arxiv.org/abs/2212.09587}{{\ttfamily arXiv:2212.09587
  [hep-lat]}}.

\bibitem{Ishii:2013ira}
{\bfseries HAL QCD} Collaboration, N.~Ishii, ``{Baryon-baryon Interactions from
  Lattice QCD},'' \href{https://dx.doi.org/10.22323/1.172.0025}{{\em PoS}
  {\bfseries CD12} (2013) 025}.

\bibitem{Aoki:2023qih}
S.~Aoki and T.~Doi, {\em {Lattice QCD and Baryon-Baryon Interactions}},
  \href{https://dx.doi.org/10.1007/978-981-15-8818-1_50-1}{pp.~1--31}.
\newblock 2023.
\newblock \href{https://arxiv.org/abs/2402.11759}{{\ttfamily arXiv:2402.11759
  [hep-lat]}}.

\bibitem{Bicudo:2012qt}
{\bfseries European Twisted Mass} Collaboration, P.~Bicudo and M.~Wagner,
  ``{Lattice QCD signal for a bottom-bottom tetraquark},''
  \href{https://dx.doi.org/10.1103/PhysRevD.87.114511}{{\em Phys. Rev. D}
  {\bfseries 87} no.~11, (2013) 114511},
  \href{https://arxiv.org/abs/1209.6274}{{\ttfamily arXiv:1209.6274 [hep-ph]}}.

\bibitem{Briceno:2017qmb}
R.~A. Briceno, J.~J. Dudek, R.~G. Edwards, and D.~J. Wilson, ``{Isoscalar
  $\pi\pi, K\overline{K}, \eta\eta$ scattering and the $\sigma, f_0, f_2$
  mesons from QCD},'' \href{https://dx.doi.org/10.1103/PhysRevD.97.054513}{{\em
  Phys. Rev. D} {\bfseries 97} no.~5, (2018) 054513},
  \href{https://arxiv.org/abs/1708.06667}{{\ttfamily arXiv:1708.06667
  [hep-lat]}}.

\bibitem{Berwein:2024ztx}
M.~Berwein, N.~Brambilla, A.~Mohapatra, and A.~Vairo, ``{Hybrids, tetraquarks,
  pentaquarks, doubly heavy baryons, and quarkonia in Born-Oppenheimer
  effective theory},''
  \href{https://dx.doi.org/10.1103/PhysRevD.110.094040}{{\em Phys. Rev. D}
  {\bfseries 110} no.~9, (2024) 094040},
  \href{https://arxiv.org/abs/2408.04719}{{\ttfamily arXiv:2408.04719
  [hep-ph]}}.

\bibitem{Nakanishi:1969ph}
N.~Nakanishi, ``{A General survey of the theory of the Bethe-Salpeter
  equation},'' \href{https://dx.doi.org/10.1143/PTPS.43.1}{{\em Prog. Theor.
  Phys. Suppl.} {\bfseries 43} (1969) 1--81}.

\end{thebibliography}

\end{document}